\documentclass[useAMS,a4paper,usenatbib,usegraphicx]{mn2e}
\usepackage{fixltx2e}          
\usepackage{txfonts}           
\usepackage{aas_macros}        
\usepackage[colorlinks,citecolor=blue,linkcolor=blue]{hyperref}
\usepackage{color}    
\usepackage{ifthen}            
\usepackage[normalem]{ulem}  
\newcommand{\lcdm}{$\Lambda$CDM}       

\newcommand{\vv}[1]{\bmath{#1}}        
\newcommand{\mat}[1]{\mathbfss{#1}}         
\newcommand{\T}{\rmn{T}}      
\newcommand{\Tr}{\rmn{Tr}}    
\newcommand{\cvv}[3]{\left(\begin{array}{l} #1 \\ #2 \\ #3 \end{array}\right)}
\newcommand{\dotprod}{\mathbf{\cdot}}

\newcommand{\dd}{\rmn{d}}             

\newcommand{\Np}{N_\rmn{p}}
\newcommand{\Msol}{\rmn{M_{\sun}}}   
\newcommand{\munit}{ \, h^{-1} \Msol }  

\newcommand{\OmegaL}{\Omega_\Lambda}
\newcommand{\OmegaMzero}{\Omega_{\rmn{M},0}}
\newcommand{\OmegaLzero}{\Omega_{\Lambda,0}}
\newcommand{\Omegabzero}{\Omega_{\rmn{b},0}}

\newcommand{\rhocrit}{ \rho_\rmn{crit} }
\newcommand{\Mpc}{\rmn{Mpc}}            
\newcommand{\kpc}{\rmn{kpc}}            
\newcommand{\kms}{\rmn{km\,s^{-1}}}     
\newcommand{\Hunit}{ \, \kms \, \Mpc^{-1}} 
\newcommand{\lunit}{ \, h^{-1} \Mpc }      
\newcommand{\klunit}{\, h^{-1} \kpc }      
\newcommand{\mpart}{m_\rmn{p}}

\newcommand{\gal}{\rmn{gal}}
\newcommand{\hlo}{\rmn{h}}
\newcommand{\img}{\rmn{img}}
\newcommand{\agal}{a_\gal}
\newcommand{\bgal}{b_\gal}

\newcommand{\vagal}{\vv{a}_\gal}
\newcommand{\vbgal}{\vv{b}_\gal}
\newcommand{\vcgal}{\vv{c}_\gal}

\newcommand{\vah}{\vv{a}_\hlo}
\newcommand{\vbh}{\vv{b}_\hlo}
\newcommand{\vch}{\vv{c}_\hlo}
\newcommand{\pr}{\rmn{pr}}
\newcommand{\vahpr}{\vv{a}_\pr} 
\newcommand{\vbhpr}{\vv{b}_\pr} 
\newcommand{\vh}{\vv{v}_\hlo}
\newcommand{\vzero}{\vv{v}_0}

\newcommand{\ux}{\vv{\hat{x}}}  

\newcommand{\uxh}{\vv{\hat{x}}_\hlo}
\newcommand{\uyh}{\vv{\hat{y}}_\hlo}
\newcommand{\uzh}{\vv{\hat{z}}_\hlo}
\newcommand{\uxgal}{\vv{\hat{x}}_\gal}
\newcommand{\uygal}{\vv{\hat{y}}_\gal}
\newcommand{\uzgal}{\vv{\hat{z}}_\gal}

\newcommand{\Msmp}{\mat{M}_\rmn{Smp}}
\newcommand{\Mrdu}{\mat{M}_\rmn{Rdu}}
\newcommand{\Msmpitr}{\mat{M}_\rmn{SmpItr}}
\newcommand{\Mrduitr}{\mat{M}_\rmn{RduItr}}

\newcommand{\figw}{84 mm} 
\newcommand{\figwbig}{100 mm} 
\newcommand{\figwthree}{57mm} 
\newcommand{\figwtwo}{65 mm} 
  
\newcommand{\fpath}{./} 

\defcitealias{bower06}{B06}
\defcitealias{delucia2007}{DLB07}
\newcommand{\dgal}{\citetalias{bower06}}
\newcommand{\mgal}{\citetalias{delucia2007}}

\newcommand{\annotate}{false}  
\newcommand{\dofigscripts}{false}
\newcommand{\drafting}{newver} 
\definecolor{dkgreen}{rgb}{0,0.5,0}
\newcommand{\pbscrapsty}[1]{\textcolor{red}{\sout{#1}}}
\newcommand{\pbnewsty}[1]{\textbf{\textcolor{dkgreen}{#1}}}  

\ifthenelse{\equal{\annotate}{true}}{
  \newcommand{\pbtodo}[1]{ [\textbf{\textcolor{red}{#1}}] }
}{
  \newcommand{\pbtodo}[1]{}
}
\ifthenelse{\equal{\dofigscripts}{true}}{
  \newcommand{\pbscripts}[1]{ [\textbf{\texttt{\textcolor{magenta}{#1}}}] }
}{
  \newcommand{\pbscripts}[1]{}
}
\ifthenelse{\equal{\drafting}{oldver}}{      
  \newcommand{\pbscrap}[1]{#1}                  
  \newcommand{\pbnew}[1]{}                      
} {\ifthenelse{\equal{\drafting}{newver}}{   
    \newcommand{\pbscrap}[1]{}                  
    \newcommand{\pbnew}[1]{#1}                  
  } {
    \newcommand{\pbscrap}[1]{\pbscrapsty{#1}}   
    \newcommand{\pbnew}[1]{\pbnewsty{#1}}       
  }
}

\title[WL \& Halo shapes]{Halo Shapes From Weak Lensing: The Impact of
  Galaxy--Halo Misalignment}
\author[P. E. Bett]{Philip Bett$^1$\thanks{Email: \pbnew{p.e.bett@physics.org}}\\
  $^{1}$Argelander-Institut f\"ur Astronomie, Universit\"at Bonn, Auf
  dem H\"ugel 71, D-53121 Bonn, Germany }

\begin{document}

\date{\today}
\pagerange{\pageref{firstpage}--\pageref{lastpage}} \pubyear{2011}
\maketitle
\label{firstpage}

\begin{abstract}
  We analyse the impact of galaxy--halo misalignment on the ability of
  weak lensing studies to constrain the shape of dark matter haloes,
  using a combination of the \emph{Millennium} dark matter $N$-body
  simulation and different semi-analytic galaxy formation models, as
  well as simpler Monte Carlo tests.  Since the distribution of
  galaxy--halo alignments is not known in detail, we test various
  alignment models, together with different methods of determining the
  halo shape.  In addition to alignment, we examine the interplay of
  halo mass and shape, and galaxy colour and morphology with the
  resulting stacked projected halo shape.  We find that only in the
  case where significant numbers of galaxy and halo minor axes are
  parallel does the stacked, projected halo axis ratio fall below
  $0.95$.  When using broader misalignment distributions, such as
  those found in recent simulations of galaxy formation, the halo
  ellipticity signal is washed out and would be extremely difficult to
  measure observationally.  It is important to note that the spread in
  stacked halo axis ratio due to theoretical unknowns (differences
  between semi-analytic models, and between alignment models) are much
  bigger than any statistical uncertainty: It is na\"ive to assume
  that, simply because \lcdm{} predicts aspherical haloes, the stacked
  projected shape will be elliptical.  In fact, there is no robust
  \lcdm{} prediction yet for this procedure, and the interpretation of any
  such elliptical halo signal from lensing in terms of physical halo
  properties will be extremely difficult.
\end{abstract}


\section{Introduction}
Dark matter haloes are irregularly-shaped virialised clumps of
collisionless matter.  In the simplest model of the Universe that is
most compatible with current observations (\lcdm), dark matter
dominates the mass budget and haloes form from the collapse and
hierarchical merging of matter in overdense regions. Galaxies form
from gas that originally clustered with the dark matter, following
baryonic processes (e.g. radiative cooling, star formation, etc) that
lead to structures with very different properties and behaviour to the
nearly-self-similar dark matter haloes.


Theoretical properties of dark matter haloes are now very well
established, following decades of research using $N$-body simulations
and advances in computing power.  This work includes characterising
the distributions and time-dependence of various properties, and
correlations between them.  The physical properties include spin
(angular momentum), shape, density profile and concentration,
phase-space density profile, clustering, and the relationship to
structures on larger and smaller scales.  Recent examples of such
studies using large-scale cosmological simulations include
\cite{2006ApJ...646..815S, 2006MNRAS.370.1422A, 2006MNRAS.367.1781A,
  2006MNRAS.367.1039H, 2007MNRAS.377...50H, 2007MNRAS.375..489H,
  2007MNRAS.381...41H, bett2007,bett2010, 2007MNRAS.378...55M,
  2008MNRAS.391.1940M, 2007MNRAS.381.1450N, 2008MNRAS.387..536G,
  2009ApJ...706..747Z, 2009MNRAS.393.1498D, 2010MNRAS.406..896B,
  2011MNRAS.411..584M, 2011MNRAS.413.1973W, 2011MNRAS.tmp..937L,
  2011arXiv1104.5130P} and the recent novel studies using principal
component analysis \citep{2011MNRAS.416.2388S, 2011MNRAS.415L..69J}.
Halo properties were recently reviewed in \cite{2011AdAst2011E...6T}.
Of particular interest in this paper are dark matter halo shapes,
which are known to have a broad distribution, with a preference for
prolateness.


Since ``dark'' matter is by definition transparent, it is very hard to
measure these properties directly using standard direct astronomical
observations.  Methods that utilise gravitational lensing however are
sensitive to the entire mass distribution, not just the radiating
baryonic component.  This has lead to gravitational lensing being
proposed as a key technique for studying halo properties
observationally \citep[see e.g. the reviews of][]{2008ARNPS..58...99H,
  2010RPPh...73h6901M, 2010GReGr..42.2177H}.

Early work on measuring halo mass distributions using weak
galaxy--galaxy lensing was performed by \cite{1993ApJ...404..441K,
  1996MNRAS.280..199W, 1996MNRAS.282..501W, 1997MNRAS.286..696S,
  1997ApJ...474...25S}.  Following these, \cite{2000ApJ...538L.113N}
proposed a technique for using weak gravitational lensing to measure
the ellipticity of haloes (see also \citealt{2000astro.ph..6281B,
  2002ASPC..283..177B}).  Consider the shear signal from weak lensing
of background (`source') galaxies, due to the mass associated with a
foreground (`lens') galaxy.  In practice, this galaxy--galaxy lensing
shear signal will be far too weak to be detectable from single lens
galaxies, so the signal from many lens systems needs to be stacked.  If
we compare the tangential shear either side of the lens galaxy image's minor
axis with that either side of its major axis, then we can obtain a
measurement of the flattening of the surrounding matter distribution.
However, if the ellipticities of halos and galaxies are not
consistently aligned, the stacking procedure will result in this
anisotropic shear signal being washed out.

This method was first used by \cite{hyg2004}, on data from the
Red-sequence Cluster Survey.  Assuming a model in which the lensing
halo and galaxy ellipticities are related through $e_\rmn{halo} = f
e_\rmn{gal}$, they found a best-fit value of $f= 0.77^{+0.18}_{-0.21}$
($68\%$ confidence level), and claimed to exclude the possibility of
spherical haloes ($f=0$) at $99.5\%$ confidence.  \cite{parker2007},
using the CFHT Legacy Survey, measured the ratio of the tangential
shears to be $0.76 \pm 0.10$, excluding spherical haloes at $\sim
2\sigma$ and implying \citep[via][]{2000astro.ph..6281B} a halo
ellipticity of $\sim 0.3$.  They also attempted to select mostly
elliptical galaxies, which resulted in a more significant detection of
ellipticity.

\cite{mandelbaum2006} performed a very thorough analysis using data
from the Sloan Digital Sky Survey (SDSS), which included photometric
redshifts and galaxy colours (not available to the other two studies).
However, they did not manage to definitively detect halo ellipticity,
although they found a hint at different alignments for different
galaxy types.  Their work showed how sensitive the results are to the
models used for interpretation: If they assumed Gaussian errors with a
power law density profile (as in \citealt{hyg2004}), they found $f = 0.1\pm
0.06$ and $f = -0.8 \pm 0.4$ for red and blue galaxies resprectively;
if they instead assumed non-Gaussian errors and a \cite{nfw97} density
profile, they instead found $f = 0.60\pm 0.38$ (reds) and $f =
-1.4^{+1.7}_{-2.0}$ (blues), where negative numbers mean
anti-alignment of mass and light.


Unambiguous detection of dark matter halo ellipticity has been seen as
an important goal, because it offers the prospect of falsifying
alternative theories of gravity, such as MOND/TeVeS
\citep{1983ApJ...270..365M, Bekenstein2004} or MOG/STVG
\citep{2006JCAP...03..004M, 2009CQGra..26h5002M}.  Such theories
suffer from being more theoretically and computationally challenging
compared to simple collisionless matter in Newtonian gravity, which
has resulted in their theoretical predictions being less developed at
present.  Nevertheless, the formalism for gravitational lensing has
been developed both for TeVeS \citep{Bekenstein2004,
  2006ApJ...636..565C} and recently for STVG
\citep{2009MNRAS.397.1885M}.  Predictions of lensing from
MOND actually predate the relativistic description from TeVeS
\citep{2001MNRAS.327..557M}, and predictions for the equivalent counterpart of
``halo'' shapes in MOND was given in \cite{2001MNRAS.326.1261M} and
\cite{2002ASPC..273..243S}.  A robust prediction from TeVeS/MOND is
that the lensing signal away from the lens galaxy should be isotropic.
Thus any detection of ellipticity -- regardless of whether it agrees
with the predictions from \lcdm{} simulations -- would falsify TeVeS.
However, this result will only be strictly true for a well-isolated
lens galaxy, which is harder to establish in practice.  The presence
of mass from nearby galaxies can produce effects which go against our
intuitive understanding of gravity, e.g. STVG violates Birkhoff's
theorem \citep{2009MNRAS.395L..25M}, and can appear to fit the Bullet
Cluster \citep{2007MNRAS.382...29B} (although in that context,
\citealt{2007MNRAS.380..911S} showed that neglecting the hydrodynamics
of the baryons is also greatly misleading).  Thus, interpreting the
results of anisotropic shear measurements, whether circular or
elliptical, should be done with caution.


The problem of galaxy--halo alignment is central to this work.  There
is, essentially, no robust prediction of the relative orientation of
galaxies within their haloes from theory or simulation.  This is not
to say that is has not been measured, but that the physical processes
involved vary significantly from simulation to simulation, and the
number of objects studied is often still small ($\la 10^2$) compared
to the large statistical samples used in observations and dark matter
simulations ($\ga 10^6$).  However, all simulation work has been
consistent in predicting a reasonably broad distribution of
galaxy--halo alignments, albeit with variation in the median
angle.  These include \cite{2002ApJ...576...21V},
\cite{2003MNRAS.346..177V}, \cite{2003ApJ...592..645Y},
\cite{2003ApJ...597...35C}, \cite{2005ApJ...628...21S},
\cite{2005ApJ...627L..17B}, \cite{2006PhRvD..74l3522G},
\cite{2009MNRAS.400...43C}, \cite{2009ApJ...702.1250R},
\cite{bett2010}, \cite{2010MNRAS.405..274H}, and \cite{deason2011}.


The qualitative impact of galaxy--halo misalignment on the method of
\cite{2000ApJ...538L.113N} is intuitive and well-known, but it has not
been considered quantitatively.  On the other hand, different models
of galaxy--halo alignment have been used for studies of the intrinsic
alignment problem in galaxy--galaxy lensing
\citep{2000MNRAS.319..649H, 2004MNRAS.347..895H, 2006MNRAS.371..750H},
and for modelling the satellite galaxy distribution when considering
cluster lensing \citep{2009ApJ...694..214O, 2009ApJ...694L..83O}.
Furthermore, variation between the predicitons of different galaxy
formation simulations and models, and even from different methods of
measuring shapes of simulated haloes, are rarely considered when
observations are compared to ``the'' theoretical prediction.  The
complex systematic problems that can affect observations and prevent
straightforward interpretation, are however very well studied
(e.g. \citealt{2000astro.ph..6281B, hyg2004, 2005MNRAS.361.1287M,
  mandelbaum2006, 2010MNRAS.407..891H}).

In this paper, we focus therefore on quantifying the impact of
galaxy--halo misalignment on stacked projected halo shapes, using a
range of different models for galaxies, halo shapes and alignment
distributions to highlight the uncertainty in the theoretical
prediction.  We do not proceed to make a direct anisotropic shear
predicion from our results, as this is already well studied
\citep[e.g.][]{2010MNRAS.407..891H}, and will only serve to reduce any
ellipticity signal.

This paper is organised as follows.  In section~\ref{s:method}, we
describe in detail the simulation and series of models we use.  This
includes the simulation and galaxy formation models
(section~\ref{s:sim}), different methods of measuring halo shapes from
simulations (\S\ref{s:shapes}), and the different alignment models we
consider (\S\ref{s:galrot}).  Section~\ref{s:mc} describes simple Monte
Carlo tests of the impact of our alignment models on distributions of
halo shapes.  We present our results in section~\ref{s:res}, as series
of axis ratios generated by stacking samples of projected halo shapes,
showing how they depend on the distributions of halo and galaxy
properties.  We discuss our conclusions in section~\ref{s:conc}.


\section{Modelling the impact of misalignment}\label{s:method}
\subsection{The simulation}\label{s:sim}
We use the original\footnote{\pbnew{A second Millennium Simulation
    (MS-II) has since been performed, using the same number of
    particles in a smaller volume; see \cite{2009MNRAS.398.1150B} for
    details.}} \emph{Millennium Simulation} (MS,
\citealt{springelMS2005}), a very large $N$-body cosmological dark
matter simulation of the large-scale structure of a \lcdm{} universe.
This simulation resolves many millions of objects at each timestep,
providing the statistical power for decribing distributions of dark
matter halo properties very percisely.  The simulation is in a
periodic box of length $500 \lunit$, populated with over $10$ billion
collisionless dark matter particles ($2160^3$), each of mass $\mpart =
8.60657\times 10^8 \munit$ and a gravitational softening length of
$5.0\klunit$.  The simulation code used was \textsc{L-Gadget-2}, a
version of the Tree-PM code \textsc{Gadget-2} \citep{gadget2} that was
specially optimised for massively parallel computations and low memory
consumption.

The MS uses a set of cosmological parameters that were chosen to be consistent
with the results of the 2dFGRS \citep{2002MNRAS.337.1068P} and WMAP-1
\citep{2003ApJS..148..175S}.  We write cosmological density
parameters as $\Omega_i(z) = \rho_i(z) / \rhocrit(z)$, in terms of the
mass density\footnote{One can write the equivalent mass-density of the
  cosmological constant $\Lambda$ as $\rho_\Lambda = \Lambda c^2/
  (8\pi G)$.} of component $i$ and the critical density $\rhocrit(z) =
3H(z)^2 / (8\pi G)$, where the Hubble parameter is $H(z)$.  For the
cosmological constant, total mass, and baryonic mass, the MS uses
values of $\OmegaLzero \equiv \OmegaL(z=0) = 0.75$, $\OmegaMzero=
0.25$, and $\Omegabzero = 0.045$.  The present-day value of the Hubble
parameter is parameterised in the standard way as $H_0=100h \Hunit$,
where $h=0.73$.  The spectral index is $n = 1.0$ and the linear-theory
mass variance in $8\lunit$ spheres at $z=0$ is given by $\sigma_8 =
0.9$.

\subsubsection{Semi-analytic models}\label{s:samodels}
Various halo and galaxy catalogues from the MS have been made publicly
available through an online
database\footnote{\url{http://gavo.mpa-garching.mpg.de/MyMillennium3/}
  and \\ \url{http://galaxy-catalogue.dur.ac.uk:8080/MyMillennium/}}
\citep{LemsonMSDB}.  They are based on two independent semi-analytic
code development programmes, that of the ICC in Durham (based on the
\textsc{Galform} model), and the MPA in Garching.  While these models
(and those of other groups) have been very successful in many regards,
no model has yet matched the full distribution of galaxy properties at
all luminosities, colours and redshifts simultaneously.  The galaxy
catalogues we shall use are the ICC model of \cite{bower06} (hereafter
\dgal), and the MPA model of \cite{delucia2007} (hereafter \mgal).
Both models are based on previous codes, incorporating new features,
and retaining/improving others.  The \dgal{} model builds on the
previous models of \cite{2000MNRAS.319..168C} and
\cite{2003ApJ...599...38B}, whereas the \mgal{} model is based on the
previous work of \cite{2000MNRAS.311..576K}, \cite{subfind2001},
\cite{2004MNRAS.349.1101D}, \cite{springelMS2005},
\cite{2006MNRAS.365...11C}, and \cite{2006MNRAS.366..499D}.  Note that
further models have been produced in subsequent work by both groups.
We have chosen to use the \dgal{} and \mgal{} models because these
versions have been very widely used, and have already been subject to
detailed model comparison work.  \cite{2009MNRAS.396.1972P} recently
performed a detailed study of the different morphological mixes
predicted by these two models, and discussed the model differences
that lead to these variations.  Further model-comparison work was
carried out in \cite{2010MNRAS.406.1533D}, concentrating on the
implementations of mergers and gas cooling.

While both models have developed from essentially the same principles
(e.g. \citealt{1991ApJ...379...52W}; see also the review of
\citealt{2006RPPh...69.3101B}), and attempt to model the same
processes, significant differences nevertheless exist in the details
of the modelling.  Different methods are used for calculating the gas
cooling rates, and they use different stellar initial mass functions
and models for attenuation by dust. Both models use the same stellar
population synthesis model, and implement feedback from stellar winds
and supernovae, injecting energy back into the gas.  Galaxy mergers
(distinct from halo mergers) and disc instabilities are treated
differently in the two models, with different triggers for starbursts.
Both models also implement the growth of black holes and feedback from
AGN in very different ways.  Finally, the models also differ in the
halo definition used, the merger tree algorithm, and the way in which
galaxy calculations are linked to the merger trees.  We refer the
reader to the papers referenced above for full details of the models.

\subsubsection{Halo Identification}
We define our haloes from the simulation particles using a multi-stage process, incorporating
information about spatial clustering, binding energy, and substructure
dynamics.  This is the so-called ``merger-tree halo'' definition
originally described in \cite{2006MNRAS.367.1039H}, to which we refer
the reader for a full description.  We summarise the main points here.

The procedure starts with the simulation particles grouped by
proximity, using the well-known Friends-of-Friends algorithm (FOF,
e.g. \citealt{defw85}), with a linking length of $b=0.2$ times the
mean interparticle separation \citep[e.g.][]{2002MNRAS.332..325P}.
Within each FOF group, self-bound substrucures are identified using
the \textsc{Subfind} program \citep{subfind2001}.  This is itself a two-stage
process, first identifying candidate substructures by finding peaks in
the density field, then performing an iterative unbinding procedure,
sucessively removing particles not gravitationally bound to the
candidate (a minimum mass of $20$ particles is imposed for
substructures).  This results in a set of FOF particle groups, each
comprising some unbound particles (``fuzz'') plus zero or more
self-bound structures, usually divided conceptually into the main body
of the halo (the most-massive substructure, MMSS), plus subhaloes.

Using the FOF/\textsc{Subfind} catalogues from different output
snapshots in the simulation, merger trees are constructed, identifying
structures in one snapshot with their progenitors and descendents in
other snapshots \citep{2006MNRAS.367.1039H}.  Our haloes are defined
using information from the merger trees as a final stage of
refinement.  Firstly, the fuzz particles are excluded, leaving the
basic halo as the set of bound structures originating from the same
FOF group.  Then subhaloes are subjected to a splitting algorithm,
allowing them to be separated off from the original halo.  This
attempts to identify subhaloes that are spatially but not dynamically
linked to the halo.  For example, a subhalo might have been linked into 
a FOF group solely by fuzz particles (now excluded), or it could simply be
flying past the main halo without yet becoming bound to it.

This halo definition, and the merger trees themselves, were originally
designed for use with the \textsc{Galform} semi-analytic model
(following \citealt{helly2003}), and its application to the MS in the
\dgal{} model.  \cite{bett2007} studied the effect of different halo
definition algorithms, comparing haloes from this method with those
from simply using FOF without refinement, and those defined by an
overdensity criterion to give a spherical halo boundry at the virial
radius.  In addition to a visual comparison (in real and velocity
space), they also compared halo spin, shapes and clustering.  In terms
of halo shapes, haloes defined by a spherical boundry were
(unsurprisingly) biased towards spherical, and the simple FOF haloes
had a much broader distribution of shapes than the merger tree haloes.
It should be noted that, while further testing of different
halo-finding algorithms is beyond the scope of this paper, the choice
of algorithm will \pbnew{affect} the results and
should be borne in mind when interpreting results here and in other
studies.

The \mgal{} model uses a slightly different halo definition, omitting
the splitting procedure outlined above.  This means that, from the
point of view of the \mgal{} galaxies, a halo consists of \emph{all}
the bound structures associated with the parent FOF halo.  These halo
catalogues therefore have slightly fewer objects than the halo
catalogues we use here (and were used in \dgal{}).  However, since
both halo catalogues are built up from the same set of
\textsc{Subfind} structures, it is straightforward to identify
galaxies from both models that are associated with the same
corresponding dark matter structure.

\subsubsection{Selecting halo--galaxy systems}
In this paper, we are interested in the possibility of measuring the
shapes of sub-cluster-mass haloes observationally.  Therefore, we
should attempt to use observational selection methods when picking
objects for study from the raw halo catalogues.  At the same time, it
is important when working with $N$-body simulations to define and
select objects for study carefully to guard against biases due to
numerical effects.

A \pbnew{technique} commonly-used \pbnew{when} selecting
haloes from simulations is to attempt to exclude unvirialised systems.
\cite{bett2007} applied a cut on the halo energy ratio (to select
haloes in `quasi-equilibrium', as an approximation to virialisation),
only accepting haloes with $|1+2T/U| \le 0.5$, where $T$ is the
kinetic and $U$ the potential energy.  While \cite{bett2007} studied
halo shapes in the MS, a large part of that paper was focused on the
halo spin parameter $\lambda$.  As originally defined
\citep{1969ApJ...155..393P, 1971A&A....11..377P}, $\lambda$ is really
only valid for isolated, virialised haloes, so this cut played an
important role in excluding invalid objects.  More generally, a
virialisation-based cut can help exclude haloes that are poorly
defined, for example those that are currently undergoing a merger.  In
this case, the boundaries of the halo itself, and thus most of its
other properties, are also poorly defined.  However, aspects of our
halo definition -- excluding unbound particles, and splitting off
dynamically separate subhaloes -- go a long way towards solving these
problems, such that the explicit cut in $|1+2T/U|$ only effects a
relatively small minority of haloes (see \citealt{bett2007}).  Since
such a cut would be very difficult to apply accurately in
observational data, we choose to not apply it here.

Another important cut usually applied to simulations is on the minimum
number of particles for a halo, to ensure that haloes are
well-resolved.  \cite{bett2007} showed that the shapes of haloes in
the MS realised with fewer than $\sim 10^3$ particles were biased away
from spherical towards prolateness.  We do not automatically apply
this cut, but we will test its impact on our results.  This is
related, in principle, to the cut in galaxy magnitude we describe
below.

The physical processes experienced by galaxies in clusters are
different to those of galaxies in lower-mass haloes.  Furthermore, the
observational techniques used to study them are also different; the
method we are concerned with here does not apply in the same way.  We therefore
exclude galaxy clusters, by applying an upper mass cut of $M <
10^{13}\munit$ (in practice, a particle-number cut at $\Np<11\,619$).
This cut is also difficult to perform observationally, but it could be
approximated by, for example,  excluding the brightest galaxies
(presuming that they are BCGs), or excluding regions where the galaxy number
density is high.  Some methods are detailed in \cite{2005ApJ...628L.101B}.

We are interested in the shape of haloes of individual galaxies. We
therefore need to maintain a 1:1 relationship between galaxies and
haloes: this means excluding satellite galaxies and subhaloes, and
restricting ourselves to central galaxies only.  (While the
  distinction between haloes and subhaloes is very important when
  analysing data from simulations, it is admittedly much harder to
  ascertain observationally.)  Since the halo definition algorithm we
use corresponds to that used in the \dgal{} model, we shall base our
analysis on that catalogue.  This means that we can simply select
galaxies from \dgal{} that have been tagged as `centrals' in the
database (i.e. using \texttt{Type = 0} in the \dgal{} database table).

Galaxies from the \mgal{} catalogue are selected by identifying the
galaxies belonging to the same \textsc{Subfind} structures as the
corresponding \dgal{} galaxies.  In most cases, these will also be
central galaxies (as the MMSS of a \dgal{} halo is likely to be the
MMSS of a \mgal{} halo).  However, sometimes a halo identified in the
\dgal{} model (and our halo catalogues) will be considered to be a
subhalo in the \mgal{} model.  This means that the associated galaxy
could have evolved significantly differently to its \dgal{}
counterpart, as (in both models) central and satellite galaxies are
treated differently.  Nevertheless, each galaxy will still be the
central galaxy of the same mass structure.  While this means that our
\mgal{} galaxy sample is \emph{not} the same as just selecting
\texttt{Type = 0} galaxies in the \mgal{} database table, it should
not have a very strong impact on the main results of this paper.
\pbnew{Indeed, since this is effectively incorporating the theoretical
  uncertainty in determining satellites and central galaxies, it helps
  to overcome some of the artificialness of the \texttt{Type = 0}
  selection, and mimic to some degree the difficulty in distinguishing
  centrals/satellites observationally.  While a full study on the
  systematic differences between \texttt{Type = 0} galaxies and field
  galaxies in general is beyond the scope of this paper, it is
  important to note that the halo mass function means that most
  galaxies ($\ga 80$ per cent) of a given brightness are centrals:
  Lower-mass haloes that can host galaxies of a given magnitude as
  centrals are always much more abundant than higher-mass haloes able
  to host them as satellites (although such galaxies are likely to
  differ systematically in other properties).  Hence, the
  \texttt{Type = 0} cut retains most galaxies of each magnitude.  }

In an effort to match observational samples as much as possible, we
select (`lens') galaxies using a cut in apparent observer-frame $r$-band
magnitude.  While similar studies with the SDSS have selected galaxies
with $r\leq 19$ \citep{mandelbaum2006}, we take our limit from the
upcoming KIDS survey, and use $r<24.3$
\citep[e.g.][]{2006aglu.progE..13K, 2010gama.conf..361K}. This is
admittedly a rather optimistic limit, but we want to avoid
handicapping our data unnecessarily.

\begin{figure}
  \centering
  \includegraphics[width=\figwtwo]{\fpath 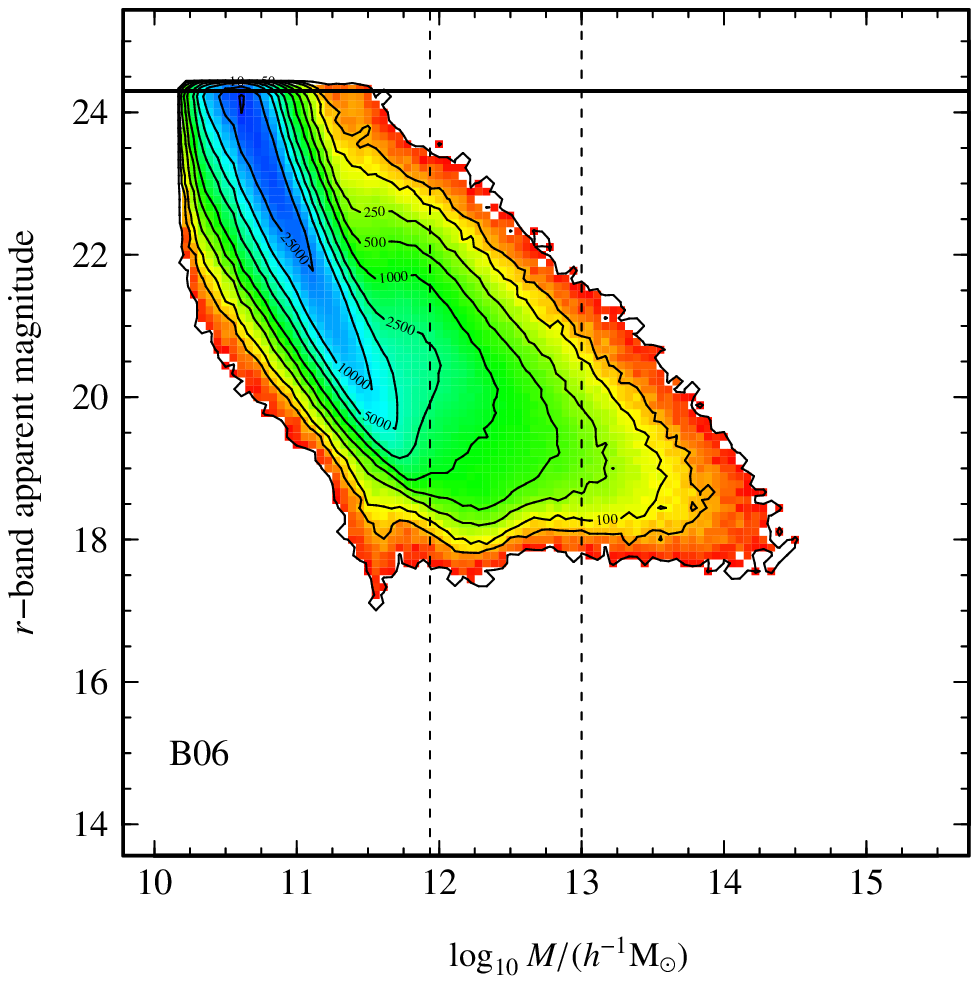}
  \includegraphics[width=\figwtwo]{\fpath 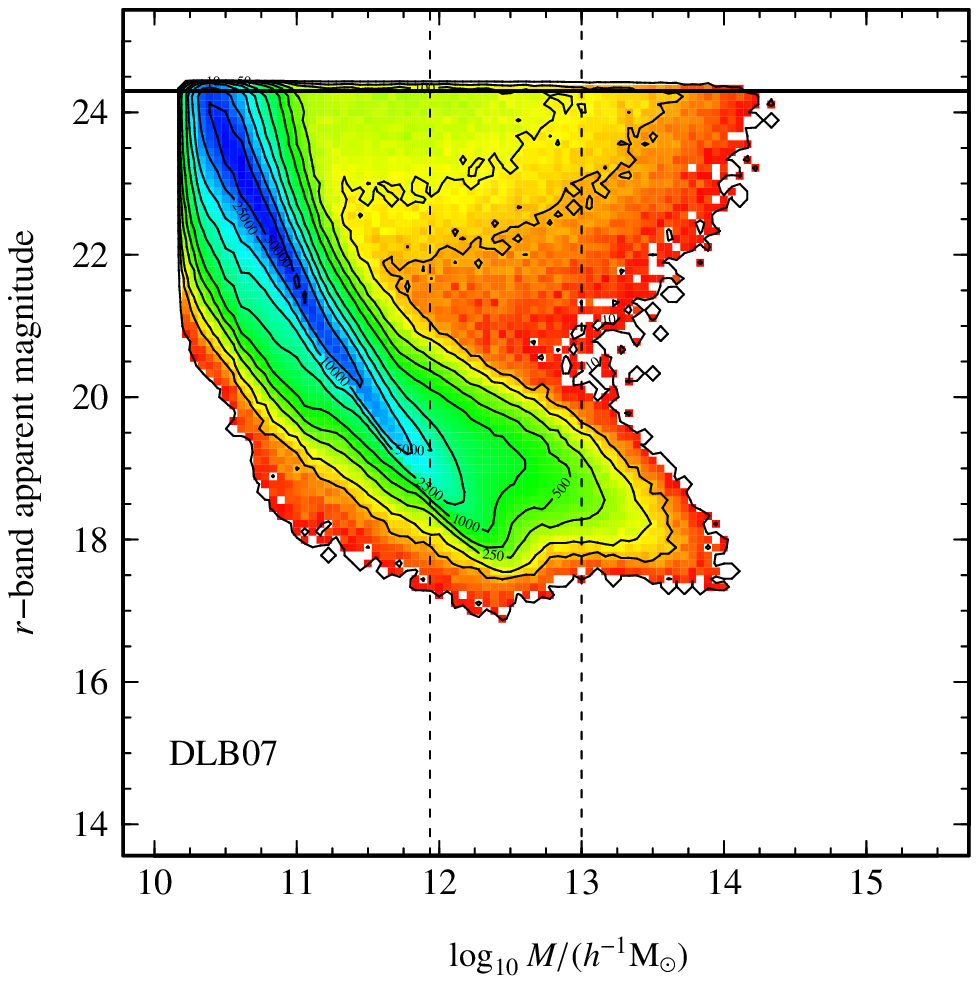}
  \caption{The joint distributions of halo mass with (observer-frame)
    $r$-band apparent magnitude, from the \dgal{} model (top) and the
    \mgal{} model (bottom).  Dashed vertical lines show the masses
    corresponding to $1000$ particles and $11619$ particles, the
    latter being the upper mass limit we will use in our analysis
    ($10^{13}\munit$).  }
  \label{f:lgmassrmag}
\end{figure}

We perform our analysis on the redshift $z=0.32$ data from the MS
(output snapshot 52), again based on the expected median redshift of
gravitational lenses in KIDS.  At this redshift, our limiting apparent
magnitude of $r_\rmn{lim} = 24.3$ translates into an (observer-frame)
absolute magnitude of $M_r - 5\log_{10}h = r_\rmn{lim} - 5\log_{10}
D(z) + 5 = -16.1$, where $D(z)$ is the luminosity distance in parsecs.
We show the joint distributions of galaxy magnitude and halo mass for
our two semi-analytic models in Fig.~\ref{f:lgmassrmag}.

While, in reality, objects would be observed over a broad range of
redshifts, it is more straightforward to use just a single snapshot
for our purposes.  This gives us many millions of objects already, so
we do not need to use other snapshots to improve our sample size.  We
do perform our analysis at other discrete redshifts however, and show
these results in Appendix \ref{s:zresults}.

In addition to galaxy magnitude, we shall also be looking at
morphology and colour as ways of selecting objects to improve halo
shape measurements.  For morphology, we use the stellar-mass
bulge-to-total ratio $B/T$.  We classify galaxies according to whether
they are bulge-dominated or disc-dominated: In particular, for
convenience we describe those with $B/T \geq 0.5$ as ``ellipticals'',
and those with $B/T < 0.5$ as ``discs''.  The $B/T$ distribution from
both semi-analytic models is
strongly bimodal: there is a very strong
peak for discs at $B/T \leq 0.005$, a much smaller but similarly
narrow peak for ellipticals at $B/T \geq 0.995$.  There is also a
significant but low-population set of intermediate-morphology
galaxies, covering $34\%$ of selected galaxies at $z\simeq 0.3$ in the
\dgal{} model, and $27\%$ for the \mgal{} model.
\cite{2009MNRAS.396.1972P} split the galaxy populations from these
models into three samples ($B/T <0.4$, $0.4\leq B/T \leq 0.6$ and $B/T
>0.6$), but for our purposes simply splitting into two samples at $B/T
= 0.5$ is sufficient.

We have tested three different measures of galaxy morphology in the
\dgal{} model: by stellar mass, by $r$-band magnitude, and by $g$-band
magnitude.  While the latter two correlate well with each other, they
can scatter somewhat when compared to morphologies determined by
stellar mass, with more galaxies appearing to have more intermediate
morphologies when determined by magnitude.  However, dividing our
galaxy population simply into two broad morphological categories means
that the vast majority of galaxies fall into the same category
regardless of the measure used.

\begin{figure*}
  \centering
  \includegraphics[width=\figwthree]{\fpath 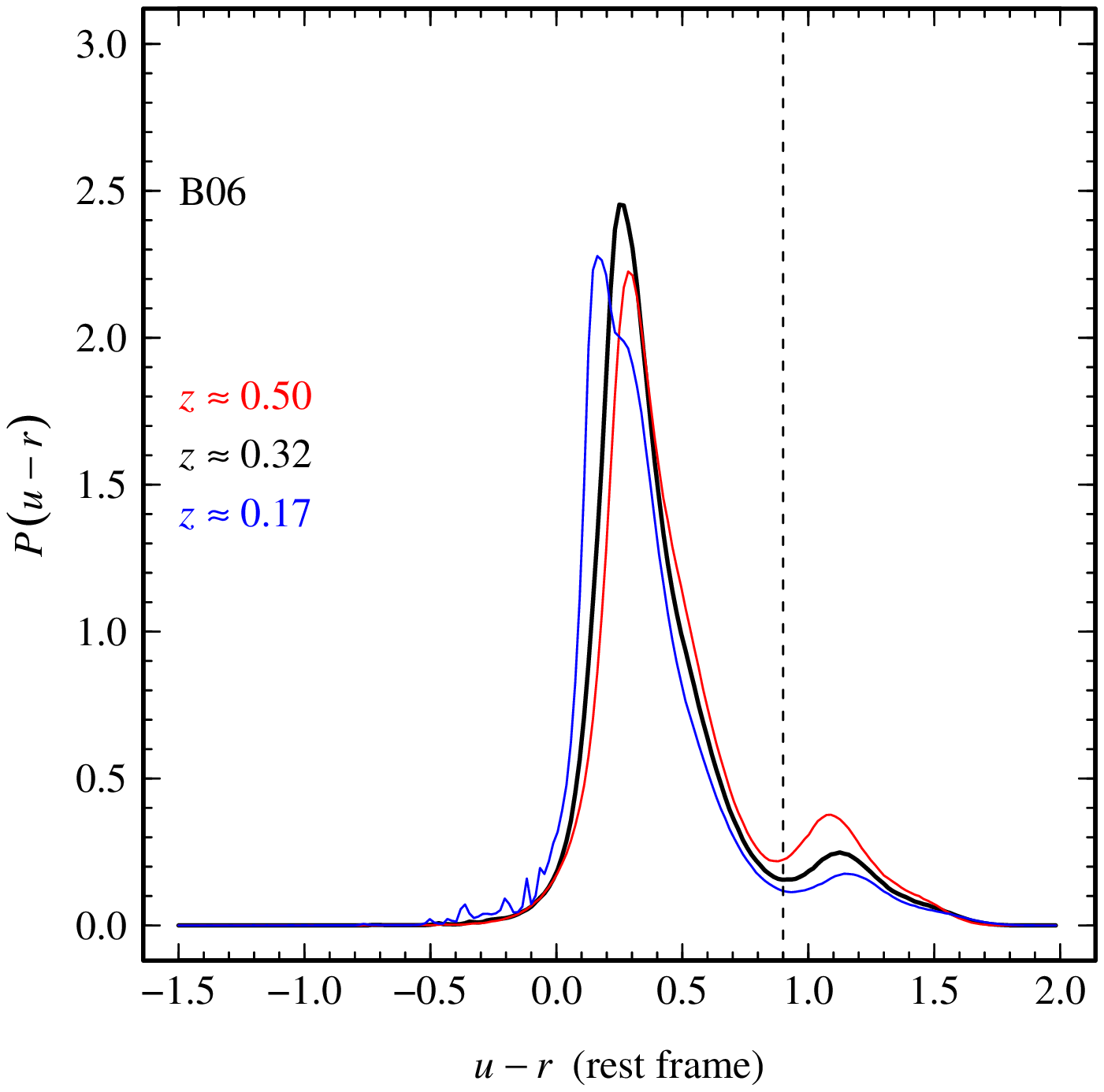} 
  \includegraphics[width=\figwthree]{\fpath 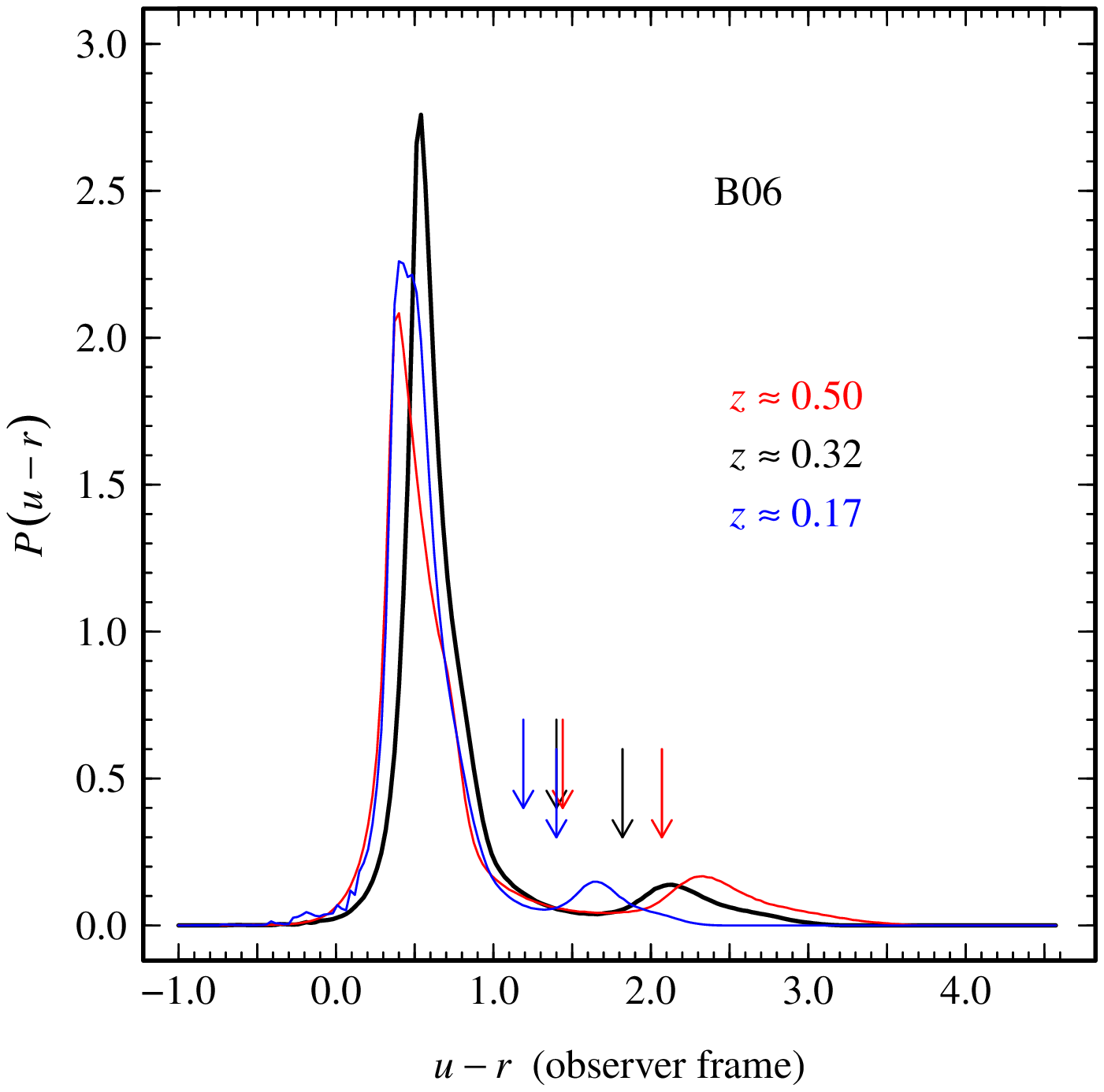} 
  \includegraphics[width=\figwthree]{\fpath 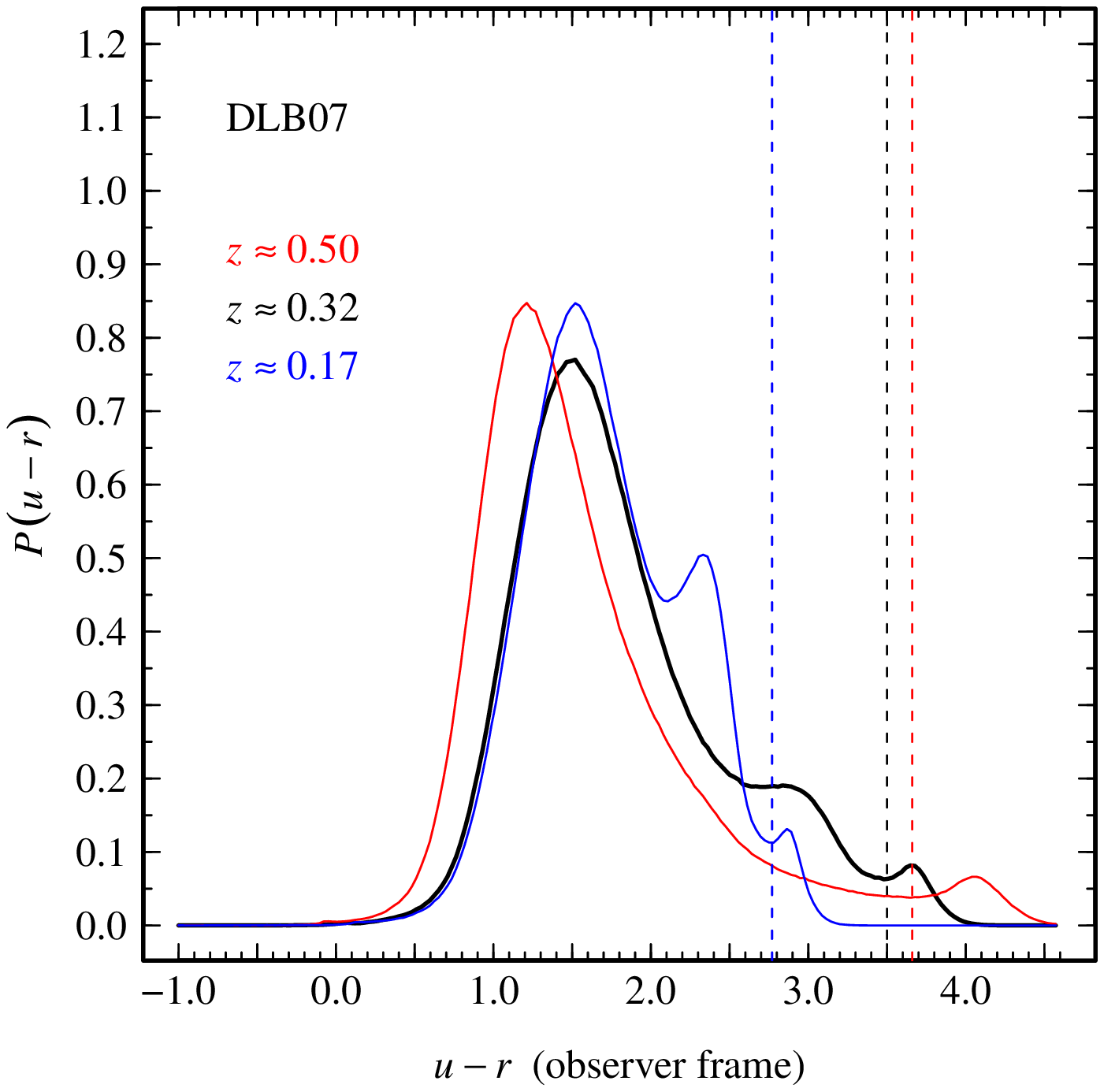} 
  \caption{The distributions of galaxy colours in the two
    semi-analytic galaxy models, at the redshift we use for our main
    analysis (black), plus the additional redshifts used in Appendix
    \ref{s:zresults} (red and blue).  Left: Rest-frame colours from
    \dgal{} with the colour-cut at $u-r = 0.9$ shown as a dashed line.
    Middle: Observer-frame colours from the \dgal{} model. Arrows
    indicate colours corresponding to the rest-frame cut, by examining
    the distribution of observer-frame colours for rest-frame $0.899 <
    u-r < 0.901$.  Upper arrows mark the medians ($1.19$ at $z=0.17$,
    $1.40$ at $z=0.32$, and $1.44$ at $z=0.50$), and lower arrows mark
    the modal colours ($1.40$ at $z=0.17$, $1.82$ at $z=0.32$, and
    $2.07$ at $z=0.50$).  Right: Same for the \mgal{} model, with the
    empirical colour-cuts marked ($2.77$ for $z=0.17$, $3.50$ for
    $z=0.32$, and $3.66$ for $z=0.50$).   }
  \label{f:coldistros}
\end{figure*}

Following \cite{2001AJ....122.1861S} (see also
\citealt{2004ApJ...600..681B}), \cite{mandelbaum2006} make the
division between ``red'' and ``blue'' galaxies at SDSS rest-frame $u-r
= 2.22$.  The semi-analytic models \pbnew{do} not
reproduce the observed colour distribution, although the colours are
easily divided into red and blue samples.  Empirically, we find that
in the \dgal{} model, we need to place that cut at rest-frame $u-r =
0.9$.  For the \mgal{} model, only the observer-frame magnitudes are
available, meaning we cannot directly compare galaxy colours at
different redshifts since a $K$-correction has not been applied.
However, examination of the colour distributions at $z\simeq 0.32$
suggests an empirical colour-cut of observer-frame $u-r = 3.5$.  We
show the colour distributions at different redshifts in
Fig.~\ref{f:coldistros}.

It is important to note that galaxy colour and
morphology are distinct, albeit related properties -- see for example
the comparison of galaxy morphologies and colour in the SDSS in
\cite{2007MNRAS.379..841B}.  We will discuss how the distributions of
colour and morphology are related for the galaxy catalogues we use, in
the Results section.

\subsection{Shapes of dark matter haloes}\label{s:shapes}
Dark matter haloes are irregularly-shaped clumps of material, defined
in principle by an envelope that demarcates either a given
mass-density threshold, or -- if one is more concerned with dynamic
properties, such as virialisation -- a threshold in phase-space
density.  In practice however, for ease \pbnew{of} both
definition and comparison between haloes, the shape of a halo is
usually characterised by the ellipsoid defined by the eigenvectors and
eigenvalues of a matrix describing the halo's internal mass
distribution.  There are many ways to measure the mass distribution
however, and different authors measure halo shapes from simulations
using different matrices.  Each of these methods have their own
advantages and disadvantages, and not all are so relevant for
observational studies.  Observations do need to be compared with
theoretical predictions from simulations however, and this can be
complicated by the variation due to the range of methods used by
theorists.  Therefore, while an in-depth comparison of different
methods of measuring halo shapes is outside the scope of this paper,
we nevertheless elect to test four different shape tensors rather than
picking just one, to highlight the variation in the theoretical
predictions.  We compare them briefly at \pbnew{the}
end of this subsection.

We describe the four halo shape algorithms we use below.  Throughout,
the tensor/vector components have indices $\{i,j\} = \{1,2,3\}$, and
the halo has $N$ particles indexed by $p$.  The particles have
positions $\vv{r}_p = (r_{p,1}, r_{p,2}, r_{p,3})^\T$ with respect to
the halo centre.

\subsubsection{The simple inertia tensor}
The inertia tensor $\mat{I}$ directly relates angular momentum
$\vv{J}$ and angular velocity $\vv{\omega}$ through $\vv{J} =
\mat{I}\vv{\omega}$, and has components 
\begin{equation}
  I_{ij} \equiv \sum^{N}_{p=1} m_p \left( \vv{r}_p^2
  \delta_{ij} - r_{p,i} r_{p,j} \right),
  \label{e:inertia}
\end{equation}
where $\delta_{ij}$ is the Kronecker delta.  Choosing a coordinate
frame in which $\mat{I}$ is diagonal (i.e. the eigenframe) is
equivalent to finding the preferred axes of rotation of the object,
i.e. the set of axes in which a torque about one does not induce a
rotation about another.  The axis directions are given by the
eigenvectors, and the eigenvalues are the moments of inertia.  The
axis lengths $a\ge b\ge c$ are given by the square root of linear
combinations of the moments of inertia per unit mass
(e.g. \citealt{bett2007}).  These principal axes define the equivalent
homogeneous ellipsoid that has the same moments of inertia -- i.e. the
same behaviour under rotations -- as the halo itself.  Axis ratios are
usually denoted $s=c/a$, $q=b/a$, and $p=c/b$.

Unless these relations to $\vv{J}$ and $\vv{\omega}$ are directly
relevant however, it is slightly simpler computationally to use the
tensor of the quadrupole moments of the mass distribution, $\mat{M}$,
which has components
\begin{equation}
  \mathcal{M}_{ij} = \sum^{N}_{p=1} m_p r_{p,i}r_{p,j}.
\end{equation}
This has the same eigenvectors as $\mat{I}$, and the eigenvalues per
unit mass give the squares of the ellipsoid axis lengths directly.
The two tensors are related through $I_{ij} = \Tr(\mat{M}) \delta_{ij}
- \mathcal{M}_{ij}$, and the quadrupole tensor is often referred to as
the inertia tensor in the literature \citep{2008gady.book.....B,
  2009ApJ...706..747Z, bett2007, bett2010}.  We shall refer to this as
the simple inertia tensor $\Msmp$ for brevity.

This tensor has been very widely used in the literature; other recent
users include \cite{2002A&A...395....1F}, \cite{2005ApJ...629..781K},
\cite{2005ApJ...618....1H}, \cite{2006ApJ...646..815S},
\cite{2006MNRAS.370.1422A},
\cite{2007MNRAS.375..489H,2007MNRAS.381...41H},
\cite{2007ApJ...671..226H} \cite{2009ApJ...702.1250R}, and
\cite{2011MNRAS.415L..69J}.

\subsubsection{The reduced inertia tensor}
A commonly-used variation on the simple inertia tensor is to
counterweight each particle by its distance from the centre, i.e. use
the tensor with components
\begin{equation}
  \label{e:redtensor}
  \mathcal{M}_{ij} = \sum^{N}_{p=1} m_p \frac{r_{p,i}r_{p,j}}{r_p^2}.
\end{equation}
This is done to remove bias due to, for example, large subhaloes
located on the outskirts of the halo \citep{1983MNRAS.202.1159G}.  In
this ``reduced'' inertia tensor (which we shall refer to as $\Mrdu$),
the particles are projected onto a unit sphere, and the shape measured
is a description of the mass distribution in different directions;
each particle contributes its mass equally.  This can provide a better
description of the ``underlying'' halo shape rather than just the
distribution of subhaloes; whether or not one considers the subhaloes
to be a distinctive aspect of the halo shape or an annoyance that
needs to be
removed depends on the study in question.  For observational studies,
the influence of the subhalo distribution is likely to be an important
part of the measurement; furthermore, the weighting would be difficult
to perform accurately.  This method has been used recently by
\cite{2005ApJ...627..647B}.

\subsubsection{The iterative simple inertia tensor}
In addition to the two preceding direct methods, iterative methods
based on the same principles are also often used.  The procedure we
use is
the following \citep[e.g.][]{1991ApJ...368..325K}:
\begin{enumerate}
\item Compute the inertia tensor $\Msmp$ using all the halo's
  particles, yielding initial axis lengths $a$, $b$, $c$. This initial
  halo has a radius $R$.
\item Select the particles within the ellipsoid just defined,
  i.e. only the particles for which the elliptical distance
  satisfies\footnote{This corresponds to keeping the major axis $a$
    constant. An alternative is to keep the volume constant, using the
    condition $\frac{r^2_{p,1}}{a^2} + \frac{r^2_{p,2}}{b^2} +
    \frac{r^2_{p,3}}{c^2} \leq 1$.}
  \begin{equation}
    \label{e:ellipr}
    \tilde{r}^2_p \equiv r^2_{p,1} + \frac{r^2_{p,2}}{q^2} + \frac{r^2_{p,3}}{s^2} 
    \leq R^2
  \end{equation}
  where we use the axis ratios $q=b/a$ and $s=c/a$.
\item Using this new particle set, recompute $\Msmp$.
\end{enumerate}
The process is deemed to converge when, after a given iteration $k$, 
\begin{eqnarray}
  \left|1-\frac{s_k}{s_{k-1}}\right| < 0.01 & \rmn{and} &
  \left|1-\frac{q_k}{q_{k-1}}\right| < 0.01.
\end{eqnarray}
The process is deemed to have failed to converge if it takes more than
$100$ steps, or the shape ellipsoid comprises fewer than $10$
particles.  We denote the resulting tensor after convergence as
$\Msmpitr$.

This procedure (or close variants of it) is often used in situations
where the set of particles comprising the object is unknown.  For
example, iterative shape-finding might be used as part of the
halo-finding algorithm, so that the resulting halo has an ellipsoidal
boundary that agrees with the measured shape exactly (within a given
tolerance), rather than the shape ellipsoid being an approximation to
the real shape of a pre-determined particle set.  Another important
usage case is when ellipsoidal density profiles are required, or just
the shape profiles themselves. Rather than use spherical bins in halo
radius, one uses the equivalent elliptical radii $\tilde{r}$, and the
axis ratios $q(\tilde{r})$ and $s(\tilde{r})$ are computed iteratively
in each bin.

Neither of these cases are relevant to us here.  Our haloes are
already defined using a more sophisticated method including particle
proximity (the FOF step), binding energy (the \textsc{Subfind} step)
and substructure dynamics (the merger tree step), and is the same
definition used for the \dgal{} galaxy model.  (A consequence of this
is that the halo shapes we measure are relatively crude approximations
to the actual isodensity surfaces.)  Furthermore, using shape profiles
is beyond the scope of the present paper.  However, an iterative
scheme is still informative, as it may give a resulting halo shape
that is more robust against numerical effects like dominance by very
few particles.  We include it here primarily to allow comparison
between different methods used in the literature.  This method has
recently been used in \cite{2006MNRAS.366.1503P},
\cite{2007MNRAS.378...55M, 2008MNRAS.391.1940M},
\cite{2011MNRAS.411..584M}, and \cite{2011ApJ...734...93L}.

\subsubsection{The iterative reduced inertia tensor}
The iteration scheme described above can be used with a reduced
inertia tensor, defined similarly to that in
equation~(\ref{e:redtensor}) \citep{1991ApJ...378..496D,
  1992ApJ...399..405W}:
\begin{equation}
  \label{e:Mrduitr}
   \mathcal{M}_{ij} = \sum^{N}_{p=1} m_p \frac{r_{p,i}r_{p,j}}{\tilde{r}_p^2}.
\end{equation}
where $\tilde{r}_p$ is the elliptical distance defined in
equation~(\ref{e:ellipr}).  We shall refer to the resulting tensor after
convergence as $\Mrduitr$.  This is, in fact, the most common way of
using the reduced inertia tensor in practice, and has been recently
used in \cite{2004ApJ...611L..73K}, \cite{2006MNRAS.367.1781A} and
\cite{2011MNRAS.tmp.1100V}.

\subsubsection{Comparison}
\begin{figure*}
  \centering\includegraphics[width=\figwbig]{\fpath 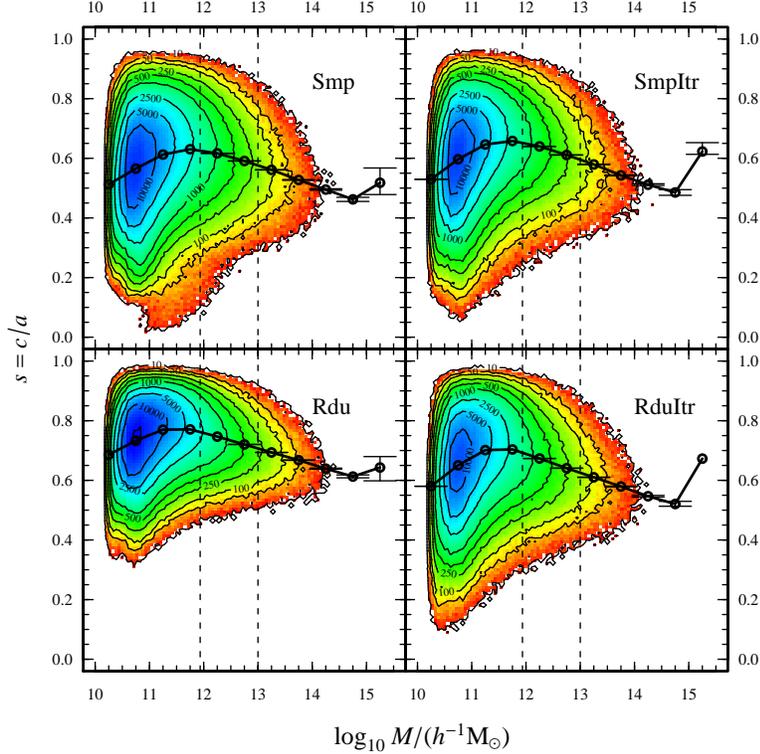}
  \caption{Halo 3-D axis ratio $s=c/a$ vs mass for the four shape
    algorithms.  Also plotted are the median values of $s$ in mass
    bins, with error bars given by equation~(\ref{e:mederr}).  Dashed
    vertical lines show the masses corresponding to $1000$ particles
    and $11619$ particles, the latter being the upper mass limit we
    use in our analysis ($10^{13}\munit$).}
  \label{f:shapemodelsmass}
\end{figure*}

To illustrate the impact that these algorithms make on the halo shape
measured in simulations, Fig.~\ref{f:shapemodelsmass} shows the
resulting axis ratios $s=c/a$ as a function of halo mass (using haloes
selected for our analysis of the \dgal{} model, i.e. at $z\simeq 0.32$,
and hosting a central galaxy from \dgal{} with $r<24.3$).  The error
bars on the medians are an estimate of their uncertainty, by analogy
with the standard error on the mean of a Gaussian:
\begin{eqnarray}
  \epsilon_+ = \frac{X_{84}-X_{50}}{\sqrt{N}}, &  \:\: & 
  \epsilon_- = \frac{X_{50}-X_{16}}{\sqrt{N}},
  \label{e:mederr}
\end{eqnarray}
where $X_i$ is the value at the $i$th percentile of the distribution
in question, made up of $N$ objects ($X_{50}$ is the median).  The
error bars only become significant at high masses, where there are
relatively few haloes in each mass bin.

We find very little difference between the results of the Simple and
the Simple Iterative shape tensors.  The reduced tensor however yields
significantly more spherical haloes at all masses, confirming the idea
that much of a halo's asphericity is due to the distributions of
subhaloes and the outer mass distribution generally.  The Iterative
Reduced shape tensor produces very similar results to $\Msmp$, but
very slightly more spherical.  Previous authors have found that, with
the advent of simulations able to resolve a significant amount of
substructure, iterative methods often failed to converge
\citep[e.g.][]{2002ApJ...574..538J, 2005ApJ...627..647B,
  2006ApJ...646..815S}.  We find that $18.0$ and $25.9$ per cent of
the selected haloes fail to converge for the $\Msmpitr$ and $\Mrduitr$
tensors respectively -- a significant amount, but a small fraction of
the population (of $7\,907\,290$ haloes).  As seen in the figure,
their loss does not bias the shape distribution significantly.

Detailed further discussion, and comparison of different shape-finding
methods, can be found in \cite{2002ApJ...574..538J},
\cite{2004IAUS..220..421S}, \cite{2005ApJ...627..647B},
\cite{2011MNRAS.tmp.1100V}, and the recent paper dedicated to the
subject by \cite{2011arXiv1107.5582Z}.

It is important to emphasize that, while all of these methods are
commonly used in the literature to measure halo shapes from
simulations, they are not all relevant for comparison with
observational studies.  In fact, it is the simple mass quadrupole
moment tensor $\Msmp$ that is the most directly related to the shear
signal from weak lensing \citep[e.g.][]{1997MNRAS.286..696S}. In the
sense that the $\Msmpitr$ is a more robust description of the same
moments of inertia as $\Msmp$, then it is also important to note if or
when it yields singnificantly different results.  However, the
``reduced'' tensors $\Mrdu$ and $\Mrduitr$, while providing very
important measures of the physical halo shape, are much less
accessible to observational tests by weak lensing.  We include them
here primarily to illustrate the systematic impact they have on the
results, to aid comparison between observational studies and future
theoretical predictions.


\subsection{Modelling the orientation of galaxies}\label{s:galrot}
The orientation of galaxies with respect to their dark matter haloes
is not tackled in current semi-analytic models of galaxy formation.
We must therefore model galaxy--halo alignment ourselves.

We consider the central galaxy within a halo, where the galaxy minor
axis $\vcgal$ is oriented in some direction $\theta$ with respect to
some characteristic halo vector $\vh$, i.e. $\vcgal\dotprod \vh =
|\vch||\vh|\cos\theta$.  In our model, we identify $\vh$ with either
the halo minor axis $\vch$, or the angular momentum $\vv{J}$. Note
that $\vch$ and $\vv{J}$ themselves have an alignment distribution,
which is not significantly correlated to other halo properties such as
shape; see e.g. \cite{bett2007} and \cite{2011MNRAS.416.2388S}.  We
can define a complete set of basis vectors for the halo ($\uxh$,
$\uyh$ \& $\uzh$), identifying the `$z$'-axis direction $\uzh$ with
that of $\vh$.  The other axes can be formed by rotations of $90\degr$
from that, following the right-hand rule.  If $\vh$ points at
\pbnew{a} polar angle $\theta_\hlo$ and azimuthal angle $\phi_\hlo$
(with respect to the simulation coordinate system, for example), then
we have
\begin{eqnarray}
  \uzh &=& \cvv{\sin\theta_\hlo \cos\phi_\hlo}
               {\sin\theta_\hlo \sin\phi_\hlo}
               {\cos\theta_\hlo},\nonumber \\
  \uxh &=& \cvv{\sin(\theta_\hlo+90\degr) \cos\phi_\hlo}
               {\sin(\theta_\hlo+90\degr) \sin\phi_\hlo}
               {\cos(\theta_\hlo+90\degr)}
         = \cvv{\phantom{-}\cos\theta_\hlo \cos\phi_\hlo}
               {\phantom{-}\cos\theta_\hlo \sin\phi_\hlo}
               {-\sin\theta_\hlo},\nonumber \\
  \uyh &=& \cvv{\sin90\degr \cos(\phi_\hlo+90\degr)}
               {\sin90\degr \sin(\phi_\hlo+90\degr)}
               {\cos90\degr}
         = \cvv{-\sin\phi_\hlo}
               {\phantom{-}\cos\phi_\hlo}
               {\phantom{-}0}.
\label{e:hbasis}
\end{eqnarray}
Note that, if $\vh\equiv \vch$, then the plane spanned by the basis vectors
$\uxh$ \& $\uyh$ is parallel to that of the halo axes $\vah$ \&
$\vbh$.  However, we do not require that e.g $\uxh$ and $\vah$
etc. are parallel, as our modelling of the galaxy orientation is based
solely on the direction of $\vh$.  The orientation of any given halo
shape with respect to its $\vh$ (and hence $\uxh$ \& $\uyh$) is fixed,
and we do not need to specify it explicitly in our modelling.

In the same way as for our halo coordinates, we let $\theta$ and
$\phi$ describe the polar coordinates giving the orientation of the
galaxy minor axis $\vcgal$, with respect to this halo reference frame.
We choose the $\theta$ and $\phi$ by randomly sampling from different
distributions.  As we see no convincing physical reason for there
being a preferred angle for $\phi$, we sample it from a uniform
distribution between $0$ and $2\pi$.  However, we test four different
models for the galaxy--halo alignment angle $\theta$, which we
describe below.

Using these two angles we can define a set of basis vectors for the
galaxy ($\uxgal$, $\uygal$, $\uzgal$) in the same way as
equations~(\ref{e:hbasis}) above.  However, if we consider the galaxy,
like the halo, to be a triaxial ellipsoid, then we need a third angle
$\xi$ to define the orientation of $\vagal$ and $\vbgal$ on the
$\uxgal$--$\uygal$ plane.  Like $\phi$, there is no convincing reason
for there to be a strongly-preferred value of $\xi$, so we again
randomly sample it from a uniform distribution over $0$--$2\pi$.  It
is important to note that $\xi$ is still significant \emph{even in the
  case of a disc galaxy with $\agal=\bgal$}.  This is because we
define our ``image plane'' later based on the galaxy's
$\vagal$--$\vcgal$ plane, so $\xi$ has a strong impact on the
orientation of the projected halo (see section~\ref{s:imgplane}).  We
give more mathematical details of the rotations involved in
implementing our orientation model in Appendix~\ref{s:rotmats}.

We now go on to describe the four models we use to provide
distributions of the galaxy--halo alignment angle $\theta$.  It should
be noted that we do not expect that the ``true'' alignment
distribution to match any of these models in detail. Rather, our
intention is that they span the possibilities of galaxy--halo
alignment, such that the impact of any given model can be easily
understood in observational terms.

\subsubsection{Parallel}
In this model, we take the characteristic halo vector to be its minor
axis ($\vh=\vch$), and set the galaxy minor axis to be perfectly
aligned with it; i.e. the angle between $\vch$ and $\vcgal$ is $\theta
= 0$.  This is the most optimistic, `best-case' scenario for attempts
to measure halo shape.

Note however that even in this case, due to our random sampling of
$\phi$, $\xi$ and the inclination of the image plane (see later), the
ellipticity of the projected shape can vary, and it can be misaligned
with respect to the galaxy.

\subsubsection{Uniform}
In this case, the orientation of the galaxy with respect to the halo
is uniformly distributed, i.e. the probability distribution of
$\cos\theta$ is flat over the range $[-1,1]$.  This is the
\emph{worst} case scenario for halo shape measurements.

\subsubsection{Fit to simulations}\label{s:fitdistro}
In the study of weak lensing with COMBO-17 data,
\cite{2004MNRAS.347..895H} used a truncated Gaussian distribution to
very roughly fit the galaxy--halo alignment from the simulations of
\cite{2002ApJ...576...21V}, which used dark matter and non-radiative gas.
In more recent years, the probability distribution for galaxy--halo
alignment has been measured in more advanced hydrodynamic simulations,
which include radiative cooling, star formation and feedback
processes.  Furthermore, we can fit them using functions more suited
to the 3-D polar angle that we are measuring.

We model the galaxy--halo alignment based on the spin--spin alignment
shown in \cite{bett2010} (their fig.~17) and
\cite{deason2011} (the top-right panel in their fig.~3), which are
based on the simulations of \cite{okamoto05} and the \textsc{Gimic}
simulations \citep{2009MNRAS.399.1773C}, respectively.  We assume that
$\vcgal$ is parallel to the galaxy spin axis
\citep{2007MNRAS.374...16L,bett2010}.  \cite{bett2010} measure the
orientation of their galaxies with respect to their parent haloes in
the galaxy formation simulation (`DMG') and also a dark matter-only
resimulation of the same initial conditions (`DMO').  We consider both
here, giving us three different datasets in total: there are $431$
galaxy--halo systems in the \cite{deason2011} data, $99$ systems from
\cite{bett2010} DMG, and $95$ from their DMO simulation.  Despite the
differences in the physics used in the different simulations, we find
that a Kolmogorov--Smirnov test fails to show a significant difference
between the three datasets at a $5\%$ significance level, i.e. they
are consistent with having been drawn from the same distibution.

\begin{figure*}
  \centering
  \includegraphics[width=\figwthree]{\fpath 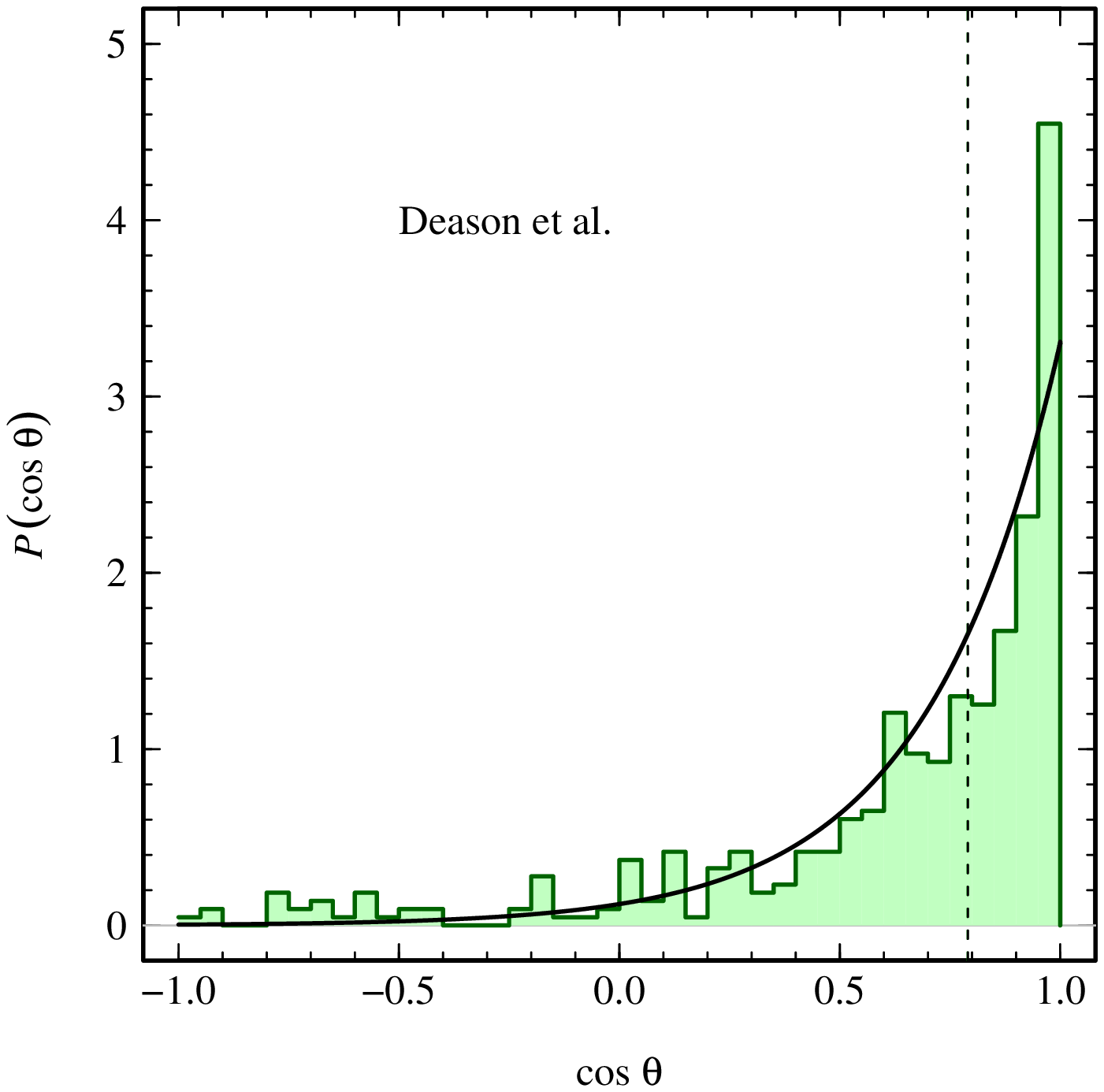} 
  \includegraphics[width=\figwthree]{\fpath 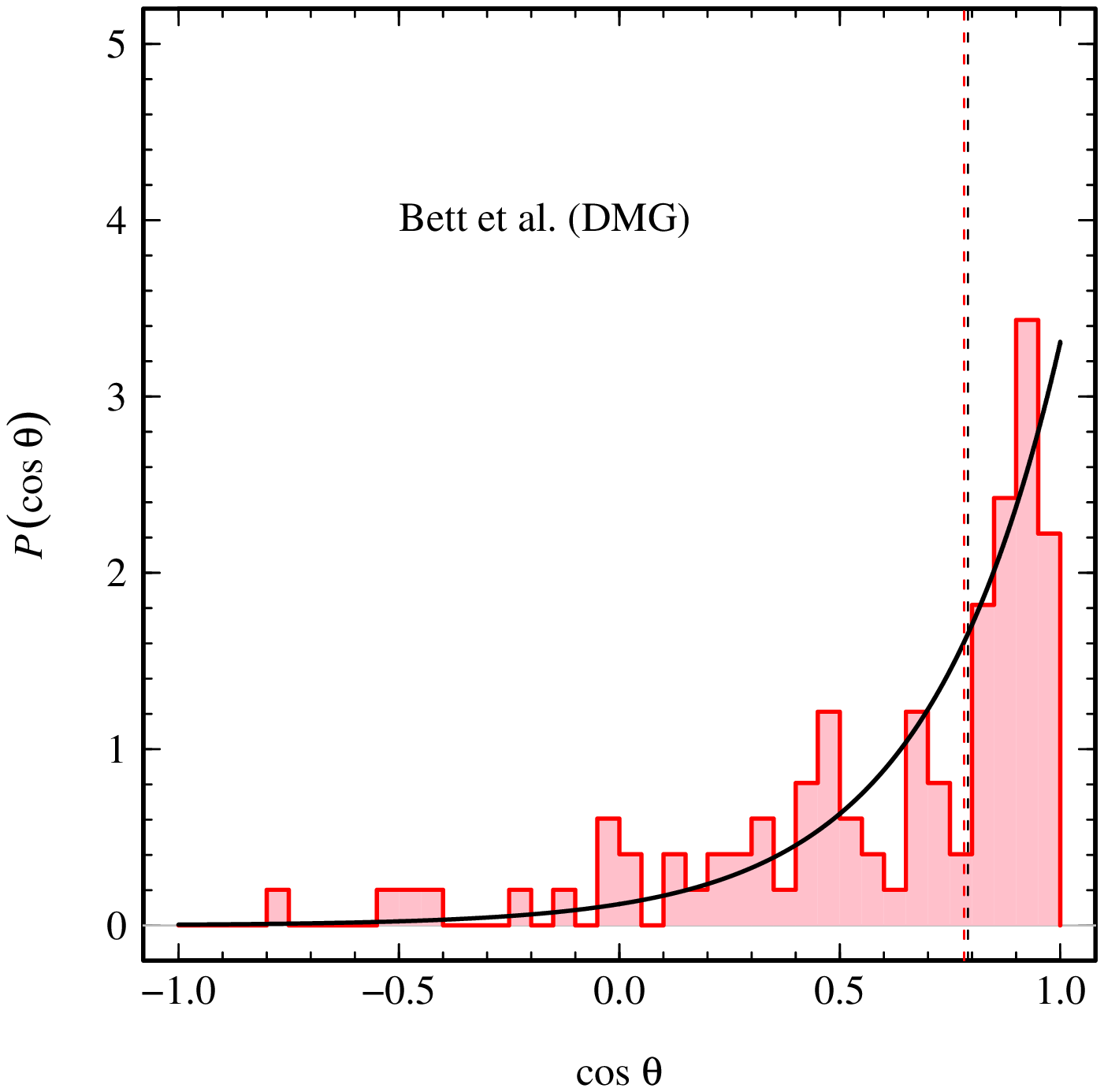} 
  \includegraphics[width=\figwthree]{\fpath 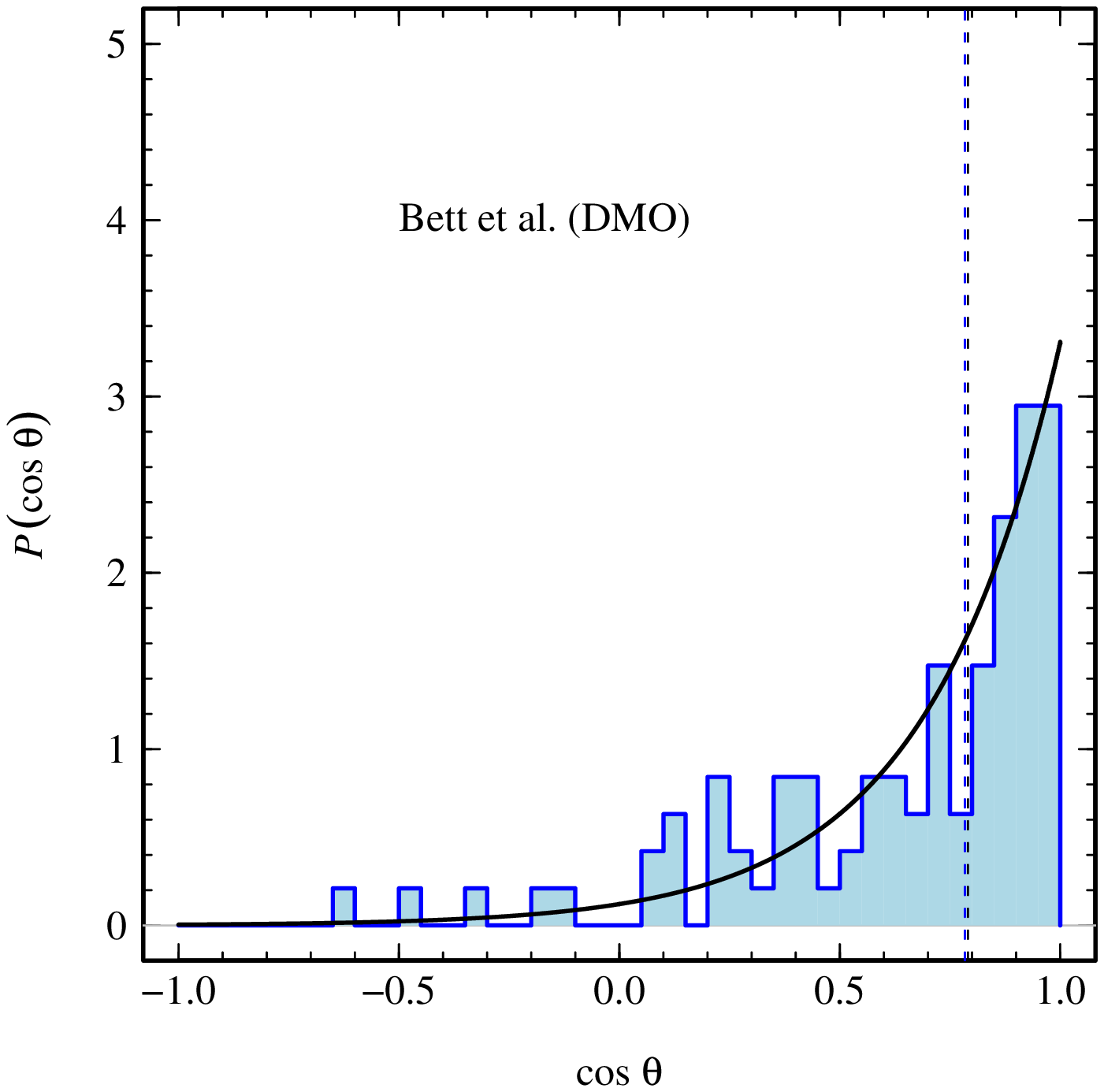} 
  \caption{Galaxy--halo alignment distributions, from
    \citealt{deason2011} (their fig.~3) and \citealt{bett2010} (their
    fig.~17).  Angles are between the halo and galaxy angular momentum
    vectors.  The black curve is the azimuthally-averaged Fisher
    distribution given by equations~(\ref{e:azavfish}) \&
    (\ref{e:fitvals}). Medians are marked with vertical dashed lines.}
  \label{f:fitdistro}
\end{figure*}

We use a \cite{fisher53} distribution averaged over the azimuthal
angle $\phi$ to characterise the alignment probability given by the
data. (We describe this distribution in more detail in Appendix
\ref{s:fishdistro}.)  The probability density function (PDF) is given
by
\begin{equation}
  \label{e:azavfish}
  P(\cos\theta) = 
  \frac{\kappa}{2 \sinh \kappa} \;
  I_0(\kappa\sin\theta\sin\theta_0) \;
  \exp{\left( \kappa\cos\theta\cos\theta_0 \right)},
\end{equation}
in terms of the ``mean'' direction $\theta_0$ and the concentration
$\kappa$, which we write in terms of the distribution width $\sigma =
1/\sqrt{\kappa}$. \pbnew{($I_0$ is the zeroth-order modified Bessel
  function of the first kind.)}  We find that sufficiently accurate
values for the mean direction and width are
\begin{eqnarray}
  \label{e:fitvals}
  \theta_0 = 0.0, & & \sigma = 0.55.
\end{eqnarray}
We show the three distributions and this fitted PDF in
Fig.~\ref{f:fitdistro}.  Note that although the preferred direction
$\theta_0=0$, the \emph{median} value of $\theta$ for this
distribution is actually $37.7\degr$.

We therefore define our third galaxy--halo alignment model as the PDF
given in equation~(\ref{e:azavfish}) with the parameters given in
equation~(\ref{e:fitvals}), using the halo angular momentum as the
reference vector $\vh = \vv{J}$.  Note that the distributions in the
Parallel and Uniform alignment models are limiting cases of
equation~(\ref{e:azavfish}), for $\sigma\rightarrow 0$ and
$\sigma\rightarrow \infty$ respectively (assuming $\vv{J}$ and $\vch$
are parallel).


\subsubsection{Split distribution}
Finally, we use a model for galaxy--halo alignment that explicitly
differentiates between different types of \pbnew{galaxies}.

The strong link between the angular momenta of gas and dark matter,
and the formation of disc galaxies, leads us to link their alignment
to the halo angular momentum.  On the other hand, ellpitical galaxies
are usually considered to have formed through mergers, which will
randomise their orientation \citep[e.g.][and references
therein]{2009MNRAS.396..696S, 2009ApJ...702.1250R,
  2011arXiv1104.0935B}.  However, the galaxy will subsequently accrete
further material, which could come from certain directions
preferentially (e.g. along filaments).  The same is true for the halo,
and we can consider the same directed merger and accretion events that
determine the halo shape to influence the galaxy shape in a similar
way.  This suggests that it is reasonable to set bulge-dominated
galaxies to be aligned to their halo.  Observational studies have
given evidence for good alignment of early-type galaxies and their
haloes (e.g. \citealt{2002sgdh.conf...62K,LensNotesSL};
\citealt{2004ApJ...613...95C}; \citealt{2008MNRAS.383..857F}; but see
\citealt{2009ApJ...694..214O}).

\cite{2000MNRAS.319..649H} introduced a simple model for galaxy--halo
alignment, in which elliptical galaxies were (implicitly) co-aligned
with their halo, and disc galaxies were aligned parallel to their
halo's angular momentum vector.  This model was also used in
conjunction with the Millennium Simulation in the work on satellite
galaxy alignments of \cite{2010ApJ...709.1321A}.
\cite{2004MNRAS.347..895H, 2006MNRAS.371..750H} extended this
alignment model to allow for a misalignment distribution around the
halo angular momentum vector, following the results from 
\cite{2002ApJ...576...21V} as already discussed.

Given the obvious similarities between our alignment models and that
of \cite{2004MNRAS.347..895H}, we construct our Split model in the
same way.  Using the stellar-mass bulge-to-total ratio $B/T$ as a
physical measure of galaxy morphology, we apply the Fitted alignment
model for galaxies with $B/T \leq 0.5$ (i.e. using
equations~(\ref{e:azavfish}) \&~(\ref{e:fitvals}) to sample $\theta$
with respect to the halo angular momentum, for disc-dominated
galaxies), and the Parallel alignment model for galaxies with $B/T >
0.5$ (i.e. $\theta=0$ with respect to the halo shape, for
bulge-dominated galaxies).

\subsection{The image plane}\label{s:imgplane}
We consider our  galaxy--halo systems as lenses, and that, when
stacked, their mass distributions will be measurable through weak
lensing of the shapes of background source galaxies.  We don't need to
actually perform the lensing itself, as we are most interested in how
galaxy--halo alignment affects the projected mass distribution; for
our purposes, the lensing process would mostly serve to add noise to
the halo ellipticity signal, making the stacked halo appear more
circular.

We assume that the observer will try to align their lens galaxies in
an image plane such that minor axis of the galaxy is parallel to the
image $y$-axis, and the galaxy major axis is parallel to the image
$x$-axis\footnote{In practice for our model, aligning with the galaxy
  \emph{intermediate} axis parallel to the image $x$ axis is
  equivalent to having $\xi=\pm 90\degr$, i.e. the uniform
  distribution of $\xi$ already accounts for the uncertainty in
  differentiating between these axes observationally.}. As galaxies
will not be exactly edge-on when observed on the sky, we have to allow
for some variation in inclination angle $\zeta$, which we define as a
rotation about the galaxy major axis (the image $x$-axis), such that
the image-plane normal vector is rotated above/below the galaxy
intermediate axis; the galaxy minor axis is rotated in front of or
behind the image plane, no longer parallel to the image $y$-axis.  We
sample $\zeta$ from a uniform distribution over the range $\pm
30\degr$.

Note that, even in the case of perfect galaxy--halo alignment in 3-D
($\theta=0$), the combination of the non-zero azimuthal angle $\phi$,
galaxy orientation $\xi$, and image-plane inclination $\zeta$ results
in misalignment between the projected galaxy and halo, and variation
in the ellipticity of the projected halo itself.  Perfect alignment in
3-D need not mean perfect alignment in projection.

\subsection{Stacking}
In observations, in order to obtain a measurable signal above the
noise from single measurements, the shear signal from many galaxy
images must be stacked, with the resulting shape being that of the net
mass distribution.  In practice, it is sensible to scale the
galaxy--halo systems to ensure that they are compared fairly, and the
signal does not become dominated by few very large systems.  This
might be done according to some spatial scale on the galaxy images, an
assumed mass content, or more directly by luminosity.  In our case,
care must be taken to use an appropriate weighting when summing
(stacking) the shape tensors of haloes. The shape tensors we use are
themselves wieghted differently: $\Msmp$ and $\Msmpitr$ scale with
halo mass and square radius (and are thus their sum is very
susceptable to dominance by high-mass haloes), whereas $\Mrdu$ and
$\Mrduitr$ just scale with halo mass.

We choose to stack halo shapes weighting by galaxy $r$-band luminosity
$L_r$, with respect to some constant reference luminosity
$L_{r,0}$. For a given shape tensor definition $\mat{M}$,
\begin{equation}
  \mat{M}^\rmn{tot} = \sum_\varsigma \frac{L_{r,0}}{L_{r,\varsigma}} 
  \mat{M}_{\varsigma}
\end{equation}
where the sum is over selected galaxy--halo systems $\varsigma$.  The
choice of $L_{r,0}$ is not important.  Since we obtain luminosities
from magnitudes, $M_{r,0} - M_r = -2.5\log_{10}(L_{r,0}/L_r)$, we
simply choose $M_{r,0} = 0$ such that $L_{0,r}/L_r = 10^{M_r/2.5}$.
Note that this will not be possible observationally if redshift
information is not available.  Instead, the strong weighting of the
shape measurement towards large haloes would be retained.  Even with
photometric redshifts, such an observational study might choose to
calculate the halo shapes in luminosity bins, rather than use
luminosity to scale the data from each lens \citep{mandelbaum2006}.

\subsection{Summary}
We have four alignment models (Parallel, Uniform, Fitted and Split)
that define the galaxy--halo alignment angle $\theta$, together with
random sampling for the azimuthal angle $\phi$, galaxy orientation
$\xi$ about its minor axis, and (over a restricted range) the image plane
inclination $\zeta$.  We also use four methods for measuring halo
shape (by the tensors $\Msmp$, $\Msmpitr$, $\Mrdu$ and $\Mrduitr$).
We are using a single algorithm for defining the haloes, and the
publicly-available results from two semi-analytic galaxy formation
models (\mgal{} and \dgal{}).

After assigning values for $\theta$, $\phi$, $\xi$, $\zeta$, we rotate
the halo shape matrix in question into the image plane.  We obtain the
eigenvalues and eigenvectors of its projection onto the image plane,
giving us the \emph{projected} halo shape axes $\vahpr$ and $\vbhpr$.
We can measure the circularity of the haloes in projection through the
axis ratio $q_\pr = b_\pr/a_\pr$.

Note that for the Parallel, Uniform and Fit alignment models there is
no link between galaxy properties and alignment.  However, since both
halo shape and galaxy properties depend on the merger history of the
halo, it is possible that halo \emph{shapes} can be correlated to
galaxy properties: in principle, one could be able to select galaxies
that preferentially have less-spherical haloes.


\section{Monte Carlo Tests}\label{s:mc}
To directly test the impact of  our alignment models and the halo
shape distribution on the resulting stacked shapes, we  perform
Monte Carlo experiments to construct a halo--galaxy sample, without
using the simulation or semi-analytic model.  

To generate a halo population, we sample the 3-D axis ratio $s$ from a
Gaussian probability distribution based on the results of
\cite{2006MNRAS.367.1781A}.  We take the standard deviation of the
Gaussian to be $\sigma_s = 0.1$, and take the mean to be
\begin{equation}
  \langle s\rangle = \alpha \left(\frac{M}{M_*}\right)^\beta
\end{equation}
where $\alpha = 0.54$, $\beta = -0.050$, and $M$ is the halo mass.
The characteristic mass $M_*(z)$ is given by
\begin{equation}
  \log_{10} \left[M_*/(\munit)\right]
  = A - B \log_{10}(1+z) - C\left(\log_{10}(1+z)\right)^2
\end{equation}
with $A=12.9$, $B=2.68$, and $C=5.96$.  For the purposes of these
tests, we take a constant halo mass $M = 10^{12}\munit$.  Using our
standard redshift of $z\simeq 0.32$, we obtain $M_* = 3.09 \times
10^{12} \munit$, leading to a distribution with a mean sphericity of
$\langle s \rangle = 0.571$.  For each value of $s$, we assign an
intermediate axis ratio of $q\equiv b/a = (1+s)/2$.

Using this halo shape distribution, we then generate samples of
projected halo shapes, each comprising $10^6$ objects.  We generate
one sample each using the Parallel and Uniform alignment models, and a
series of samples based on the Fitted model. In the latter case, we
choose a different value of the alignment distribution width $\sigma$
for each sample.  We do not model the halo angular momentum, and
instead take the alignment distribution to always be with
respect to the halo shape.  We retain the variability in image plane
alignment of $\pm 30\degr$.

We stack these projected haloes directly, giving $\mat{M}_\rmn{tot} =
\sum_\varsigma \mat{M}_\varsigma$;   since the haloes are all
the same size, we need not (and cannot!) weight by galaxy luminosity.

\begin{figure}
  \centering\includegraphics[width=\figw]{\fpath 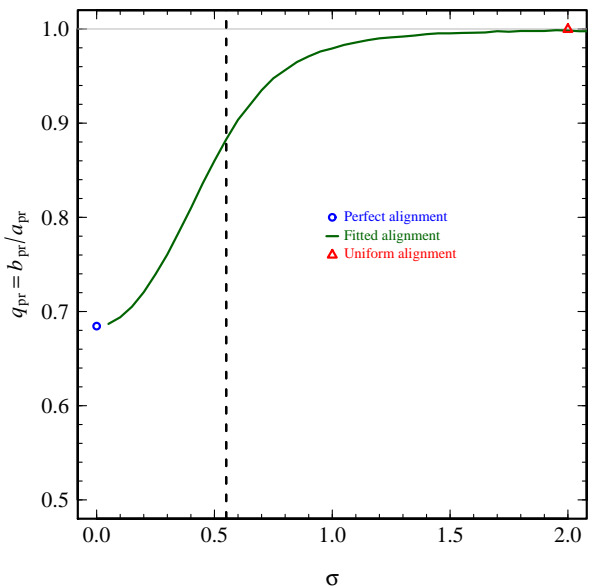} 
  \caption{Shape of stacked, projected haloes generated by sampling
    their 3-D shapes from a Gaussian, then sampling their orientation
    with respect to an image plane in the usual way; see text for
    details.  The continuous line joins the results from alignment
    distributions with a range of widths $\sigma$ (see
    equation~\ref{e:azavfish}), with the additional points (the
    circle and triangle) representing the limiting cases of
    $\sigma\rightarrow 0$ and $\sigma \rightarrow \infty$
    respectively. }
  \label{f:MCstacks}
\end{figure}

The results, showing how the resulting stacked halo shape depends on
the alignment distribution width, are shown in Fig.~\ref{f:MCstacks}.
The stacked halo shape quickly changes from $q_\pr \approx 0.68$ for
Parallel alignment, through $q_\pr >0.9$ for $\sigma \ga 0.6$, and
converging to the result from the Uniform distribution by $\sigma
\approx 2$.

\begin{figure}
  \centering\includegraphics[width=\figw]{\fpath 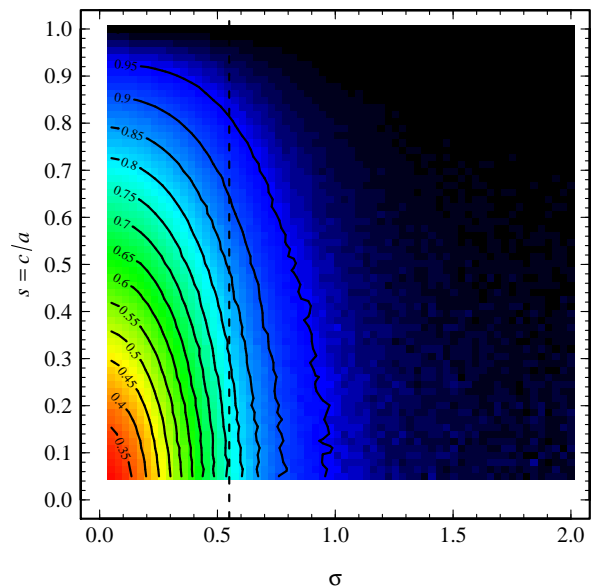}
  \caption{The contours and shading show the stacked axis ratio
    $q_\pr$ from random-sampling halo orientations in the usual
    manner, from distributions with width $\sigma$, and haloes with
    3-D axis ratio $s$.  The lowest contour (in the red region) is at
    $q_\pr = 0.35$; the other contours increase in steps of
    $0.05$.  }
  \label{f:MCstacksfixeds}
\end{figure}

We have also investigated the joint impact of the original halo shape
distribution and the alignment distribution width.  For this, we did
not sample halo shapes from a Gaussian, but instead set them all to a
fixed value $s$.  The orientation distributions were randomly sampled
in the same way as before, for a grid of values of $s$ and $\sigma$.
The results are shown in Fig.~\ref{f:MCstacksfixeds}.  It shows that,
as expected, the sphericity of the halo population is largely
immaterial, unless the alignment distribution has $\sigma \la
0.5$. Even in that case, one needs a strongly aspherical shape
distribution, with $s\la 0.3$ in order to get a stacked shape of $q_\pr
\la 0.8$.

These Monte Carlo tests have shown quantitatively the sensitivity of
the stacked halo shape on the form of the galaxy--halo alignment
distribution.  Thus we expect that, if galaxies are aligned randomly
in their haloes, or even if they are aligned as found in recent
hydrodynamic simulations, then the stacked halo shape will be $\ga
0.9$.


\section{Results}\label{s:res}
\begin{figure*}
  \centering\includegraphics[width=\figwbig]{\fpath 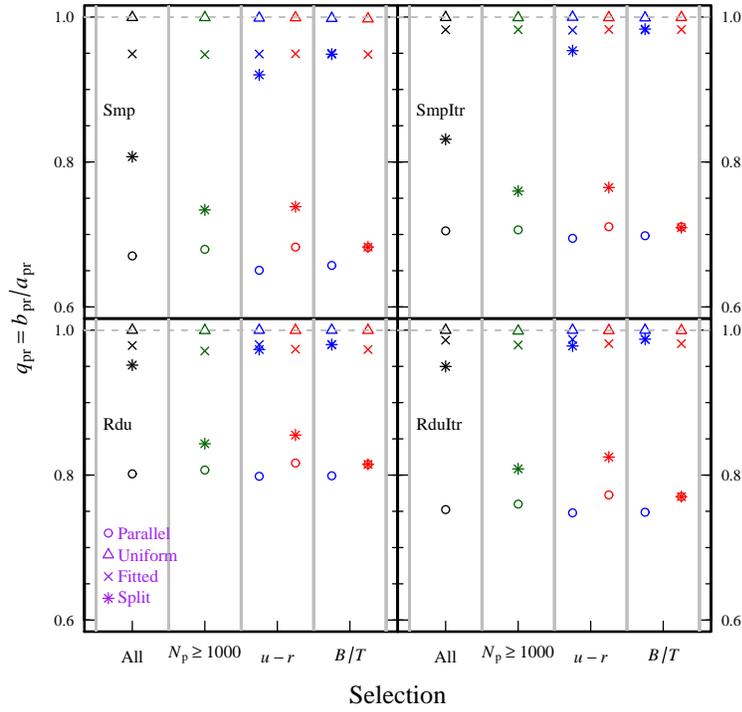}
  \caption{Axis ratios of the resulting projected shape tensors
    computed from stacking haloes selected according to different
    criteria.  This plot uses our standard redshift of $z\simeq 0.32$
    and the \dgal{} semi-analytic model.  Each of the four panels
    gives the results from using a different halo shape algorithm, and
    each symbol type gives the result from different galaxy--halo
    alignment models.  The first column in each panel (``All'') gives
    the results for all selected haloes with $M<10^{13}\munit$.  The
    next column adds a restriction at low masses, excluding haloes
    comprising $<1000$ particles.  The third and fourth columns split
    the halo population from the ``All'' column by colour and
    morphology respectively, with the blue symbols representing the
    blue/disc case, and red symbols representing the red/elliptical
    case. The colour cut is made at rest-frame $u-r = 0.9$, and the
    morphology cut is made at $B/T = 0.5$.}
  \label{f:stcksDGals}
\end{figure*}

\begin{figure*}
  \centering\includegraphics[width=\figwbig]{\fpath 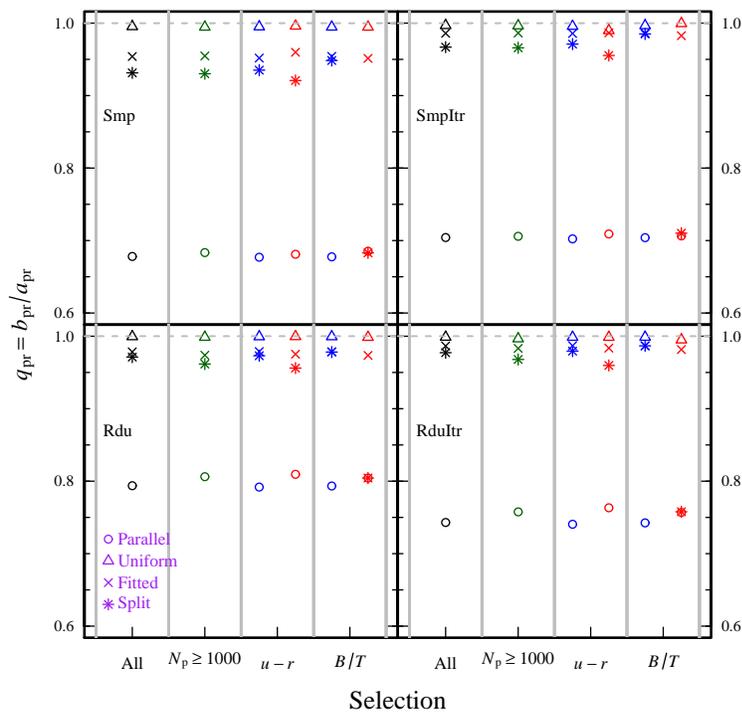}
  \caption{As Fig.~\ref{f:stcksDGals}, but using the \mgal{}
    semi-analytic model.  The colour-cut is therefore made at
    observer-frame $u-r = 3.5$.  
}
  \label{f:stcksMGals}
\end{figure*}

The results for the axis ratios $q_\pr = b_\pr/a_\pr$ of the stacked projected
halo shapes are shown in Fig.~\ref{f:stcksDGals} for the \dgal{}
model, and Fig.~\ref{f:stcksMGals} for the \mgal{} model.  Each point
represents the result for a given combination of models, with the
different columns showing the effect of different selection criteria.
We now go on to examine these results in detail.

\begin{table*}
  \begin{minipage}{126mm} 
    \caption{The number of galaxy--halo systems for the different
      models and selections in Figs.~\ref{f:stcksDGals}
      \&~\ref{f:stcksMGals}.}
    \label{t:stacknumbers}
    \begin{tabular}{ccrrrrrr}\hline
      Model   & Shapes           & All      & $\Np\ge 1000$& Blue       & Red      & Disc        & Elliptical\\
      \hline
              & $\Msmp$, $\Mrdu$ &  7\,866\,537 & 443\,053 &  7\,064\,928 & 801\,609 &  6\,890\,492 & 976\,045 \\
      \dgal{} & $\Msmpitr$       &  6\,454\,412 & 320\,280 &  5\,869\,870 & 584\,542 &  5\,738\,275 & 716\,137 \\
              & $\Mrduitr$       &  5\,830\,433 & 318\,931 &  5\,249\,670 & 580\,763 &  5\,116\,628 & 713\,805 \\
              \hline
              & $\Msmp$, $\Mrdu$ & 10\,710\,174 & 442\,862 & 10\,460\,421 & 249\,753 & 10\,241\,493 & 468\,681 \\
      \mgal{} & $\Msmpitr$       &  9\,034\,263 & 320\,177 &  8\,842\,123 & 192\,140 &  8\,660\,850 & 373\,413 \\
              & $\Mrduitr$       &  7\,887\,973 & 318\,797 &  7\,707\,025 & 180\,948 &  7\,542\,620 & 345\,353 \\
              \hline
    \end{tabular}
  \end{minipage}
\end{table*}

A quick glance confirms that the primary factor in determining the
measured stacked halo shape is the galaxy--halo alignment
distribution.  When the Uniform model is used (triangles in the
plots), the stacking process washes out any intrinsic halo
ellipticity, and the stacked halo is circular.  The maximum deviation
from circularity comes when the Parallel alignment model is applied,
as this allows the maximal contribution from all haloes towards the
final shape.

In the Parallel case, there are significant differences caused by the
different halo-shape algorithms.  When $\Msmp$ is used, the result is
furthest from circular, with larger axis ratios generated when $\Mrdu$
is used.  The iterative methods give moderated values of these
extremes: using $\Msmpitr$ yields slightly more circular haloes than
$\Msmp$, and using $\Mrduitr$ yields slightly less circular haloes
than $\Mrdu$.  This difference is due to the different implicit
weighting that these methods give to haloes when stacking.  The
reduced tensors have their dependence on halo size (radius) scaled
out, so that haloes contribute proportionally to their mass (which we
then reduce by counterweighting by luminosity).  The simple inertia
tensors however retain the additional square-radius dependence. This
means that the stacked halo results are much more strongly dominated
by high-mass objects in the simple inertia tensor case, but are more
evenly weighted in the reduced case.  How this effects the results
depends on how the intrinsic halo shape distribution varies with mass
for the different algorithms, which we showed earlier in
Fig.~\ref{f:shapemodelsmass}: higher mass haloes tend to be less
spherical. 

\pbnew{A numerical artefact, present in $N$-body
  simulations such as the MS, is that haloes consisting of a small
  number of particles tend appear systematically less spherical than
  those with many particles.  A lower limit of around 300 particles
  was suggested for the MS in \cite{bett2007} to ensure accurate halo
  shapes.}  In Figs.~\ref{f:stcksDGals} and \ref{f:stcksMGals}, we
compare the results from all haloes and those with at least $1000$
particles (in all cases, our upper mass limit of $M<10^{13}\munit$
applies).  We see that excluding the low-mass haloes makes only a very
small difference to the stacked result: For the $\Msmp$ and $\Msmpitr$
algorithms, high-mass haloes dominate the stacking anyway, and for
$\Mrdu$ and $\Mrduitr$ the fact that the numerical biasing at low
masses is in the same direction as the natural trend going to high
masses \pbnew{leaves} negligible net effect.

If, without numerical constraints, haloes in fact continue to get more
spherical towards lower masses (as suggested by
\citealt{2008MNRAS.391.1940M} and \citealt{2011MNRAS.411..584M}), then
the effect of retaining lower masses in the stacking would be more
important: when using the reduced inertia tensor, the result for
``all'' haloes in our figure would be more circular.

\subsection{Split alignment and the morphological mix}
The Split alignment model shows the greatest variation among the
models and selections tested.  Since in this model the galaxy--halo
alignment depends explicitly on galaxy morphology, the stacked results
when selecting by morphology are entirely predictable: For elliptical
galaxies, the result mirrors that from the Parallel alignment model,
whereas for disc galaxies it mirrors that of the Fitted alignment
model.  For the other selections, the result depends on how the
distribution of galaxy morphologies relates to the quantity used for
selection.

Even the data for ``All'' systems shows significant variation between
the \dgal{} and \mgal{} models, and for different shape algorithms.
Furthermore, excluding low particle-number systems has a significant
impact in the \dgal{} model, but not in the \mgal{} model.  We
therefore need to examine how the galactic morphological mix varies
with halo mass in the two models.

\begin{figure*}
  \centering
  \includegraphics[width=\figwthree]{\fpath 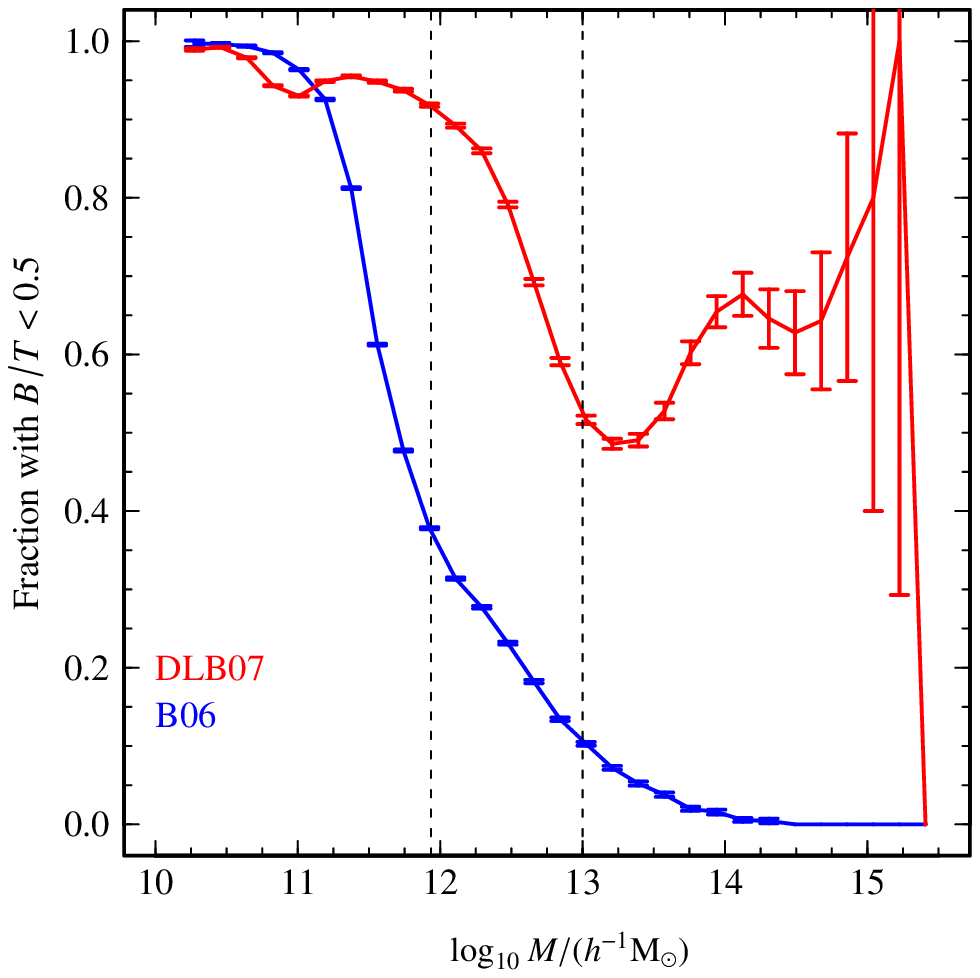}
  \includegraphics[width=\figwthree]{\fpath 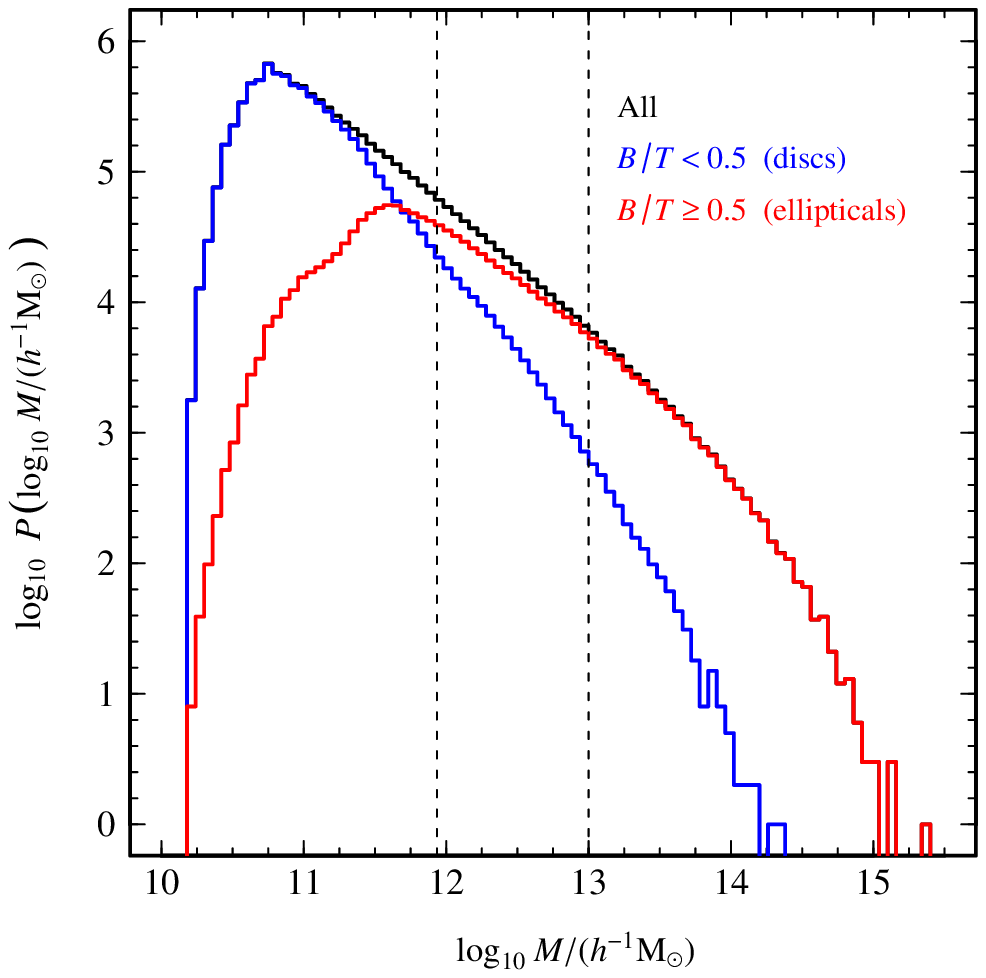}
  \includegraphics[width=\figwthree]{\fpath 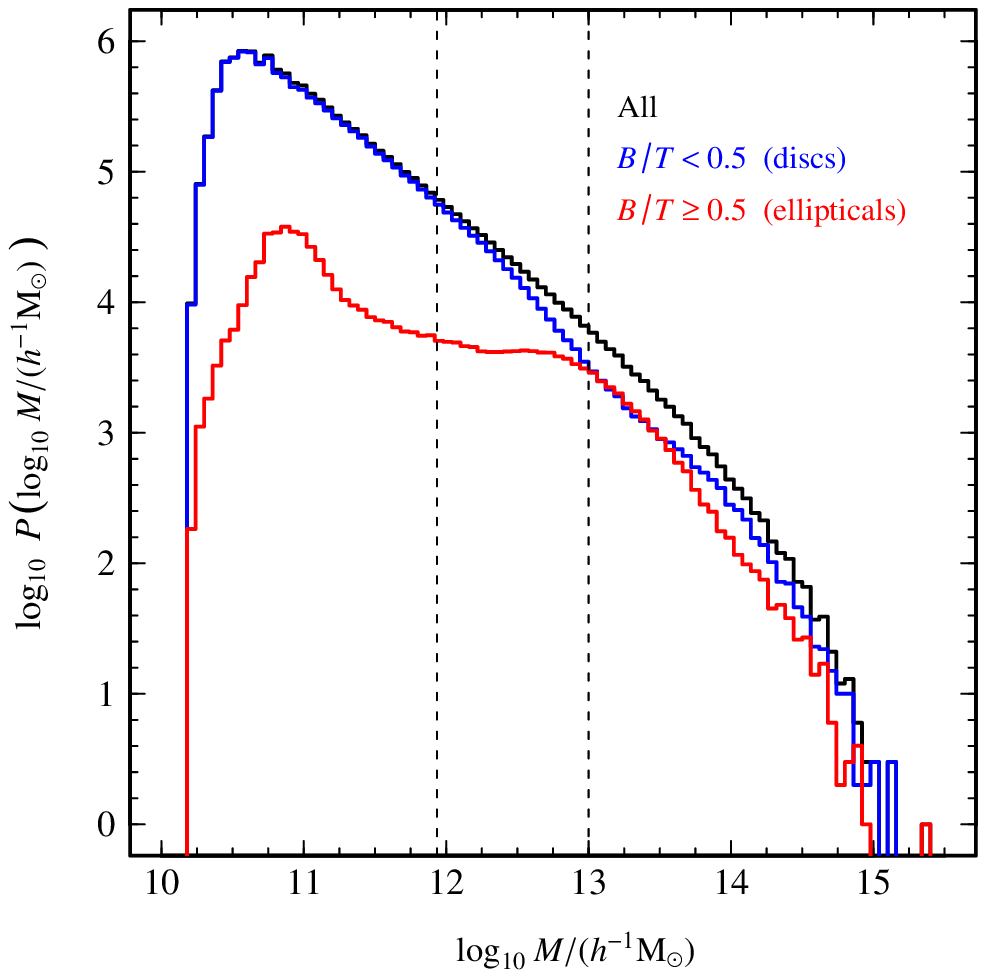}
  \caption{Distribution of galaxy morphologies as a function of halo
    mass, for the \dgal{} and \mgal{} semi-analytic models.  The left
    panel shows the proportion of galaxies at each mass that have
    $B/T<0.5$ (i.e. discs), with Poisson error bars.  The middle and
    right panels show the (normalised) histograms of the different
    galaxy samples. Note that the plots extend beyond the upper mass
    limit of $10^{13}\munit$ that we use in our analysis, and below
    the lower limit of $1000$ particles that we consider; both limits
    are marked with vertical dashed lines.}
  \label{f:morphomass}
\end{figure*}

This is shown in Fig.~\ref{f:morphomass}.  We can clearly see that,
while both semi-analytic models are dominated by disc galaxies at low
halo masses, they show very different bahaviour for higer-mass haloes.
In the \dgal{} model, the proportion of galaxies that are discs falls
rapidly with increasing halo mass, such that the galaxy population at
high masses is dominated by elliptical galaxies.  In the \mgal{} model
however, the population remains dominated by disc galaxies for roughly
another decade in mass, and only drops to a roughly even spread
between discs and ellipticals. (Note that \cite{2009MNRAS.396.1972P} \pbnew{have studied in detail} the
origin and evolution of galaxy morphologies in the \dgal{} and \mgal{}
models.)

This is reflected in the results for the stacked halo shape.  When the
$\Msmp$ tensor is used, the stacked shape is weighted towards the
high-mass haloes.  In the \dgal{} model, this means that the majority
of strongly-contributing haloes host elliptical central galaxies, with
the Parallel alignment model. There is still a significant number of
disc galaxies however, and their Fitted alignment model means that the
net stacked shape is more circular than if the Parallel alignment
model was used alone.  In the \mgal{} model, the galaxy morphologies
are even more mixed, with the additional misalignment resulting in a
more circular stacked result.

On the other hand, if a reduced shape tensor is used, then in all
cases the haloes are weighted more equally.  Those dominant by number
are at low masses, which are dominated by disc galaxies, using the
Fitted alignment distribution.  Because the steep drop in the fraction
of disc-dominated galaxies happens at lower masses in the \dgal{}
model, that model has fewer disc galaxies and results in a less
circular stacked halo shape.

\begin{figure}
  \centering
  \includegraphics[width=\figwtwo]{\fpath 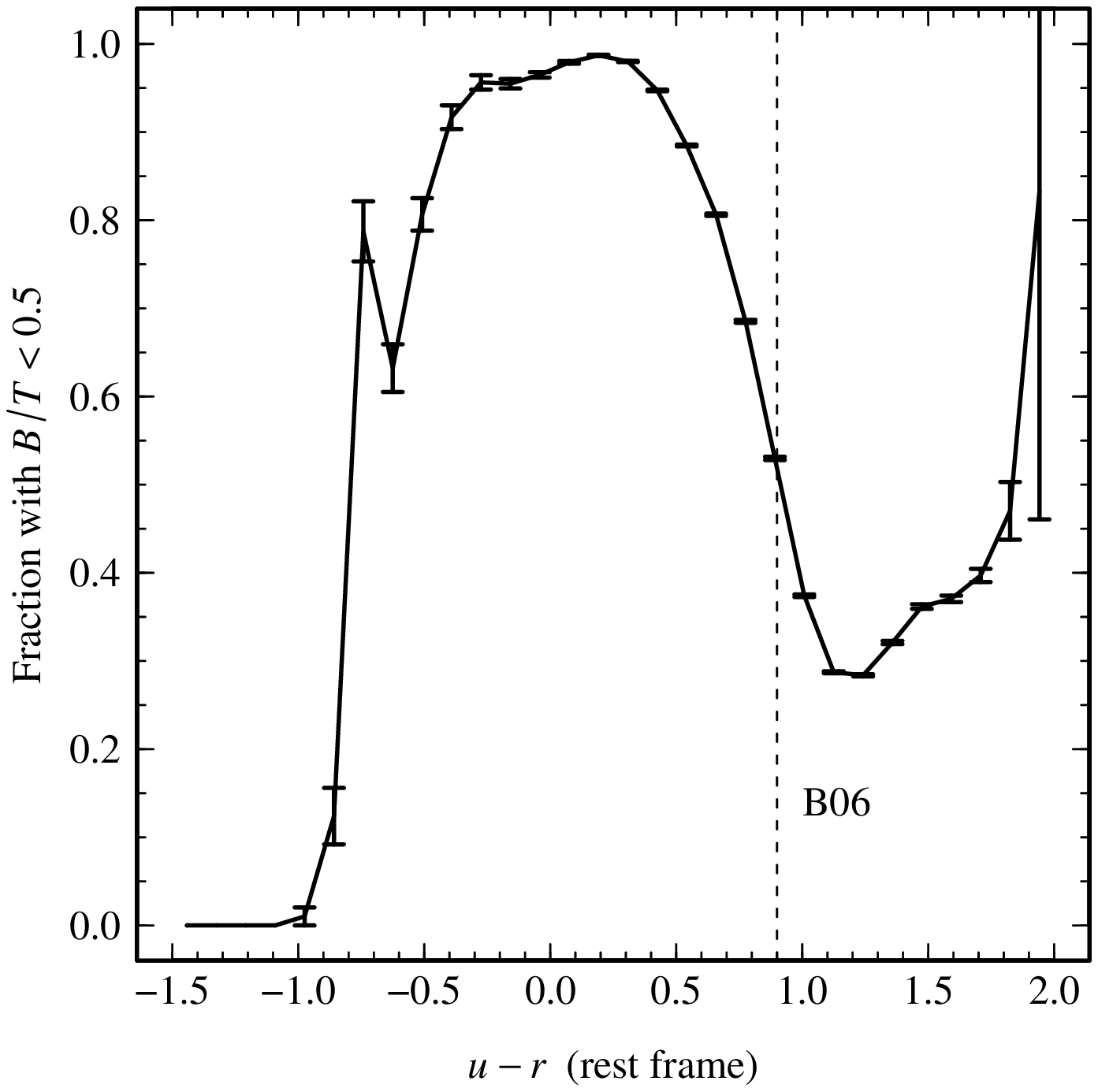}
  \includegraphics[width=\figwtwo]{\fpath 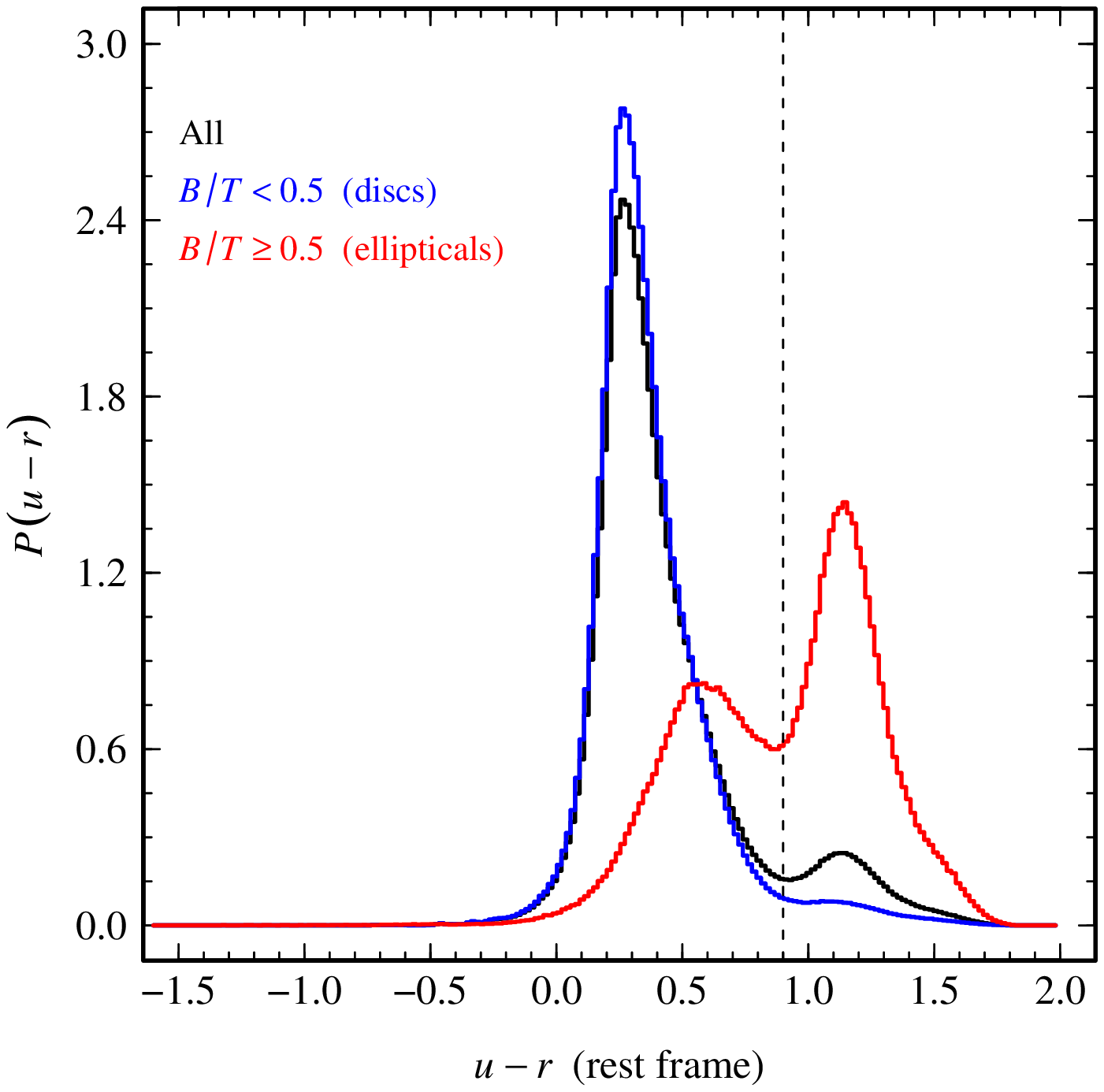}
  \caption{Distribution of galaxy morphologies as a function of
    colour, for the \dgal{} model.  The top panel gives the fraction
    at each colour that are disc-dominated, with Poisson error bars.
    The bottom panel shows colour histograms for each population
    sample.  The colour cut at rest-frame $u-r=0.9$ is marked with a
    vertical dashed line. 
}
  \label{f:morphcolD}
\end{figure}

\begin{figure}
  \centering
  \includegraphics[width=\figwtwo]{\fpath 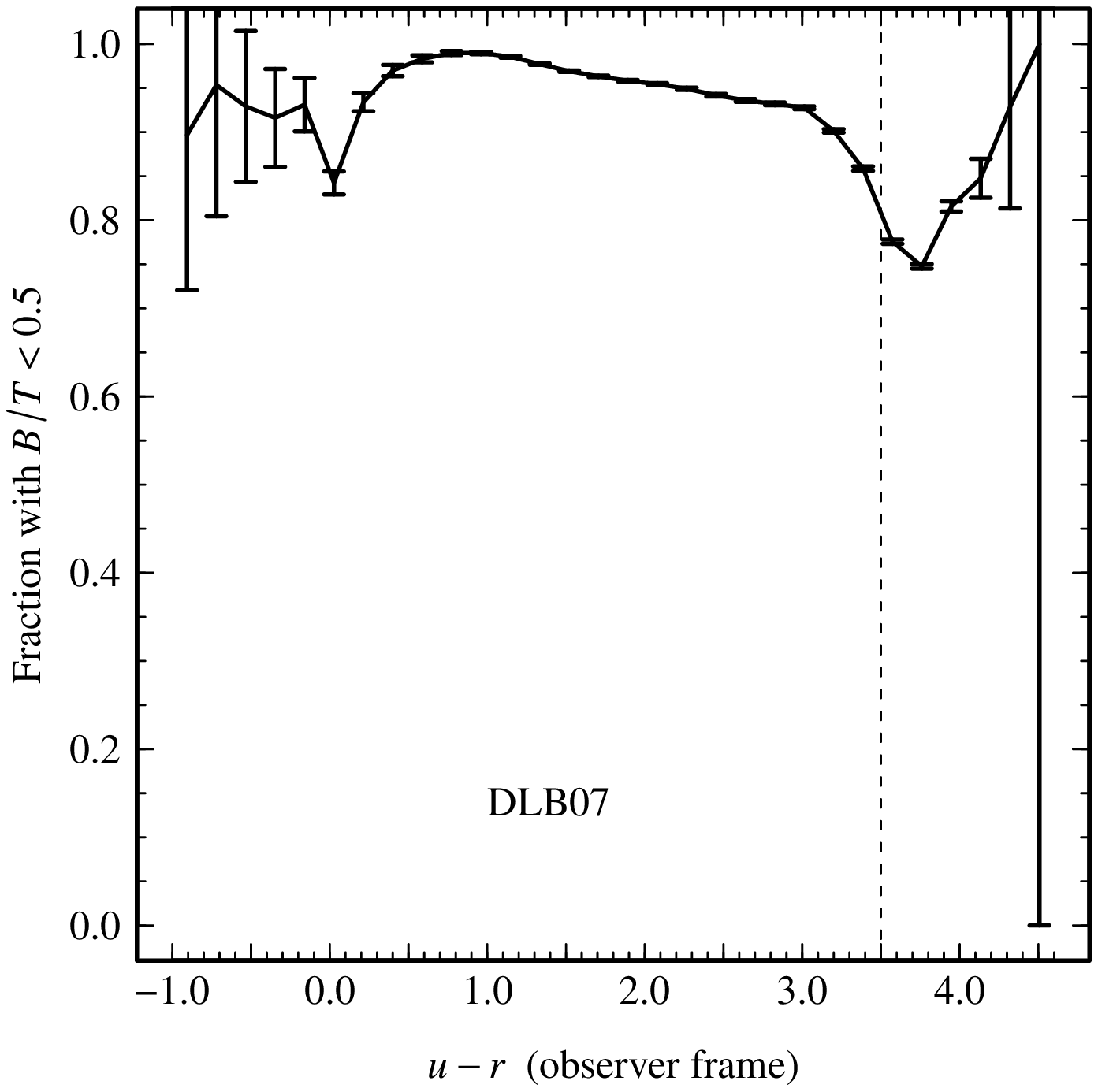}
  \includegraphics[width=\figwtwo]{\fpath 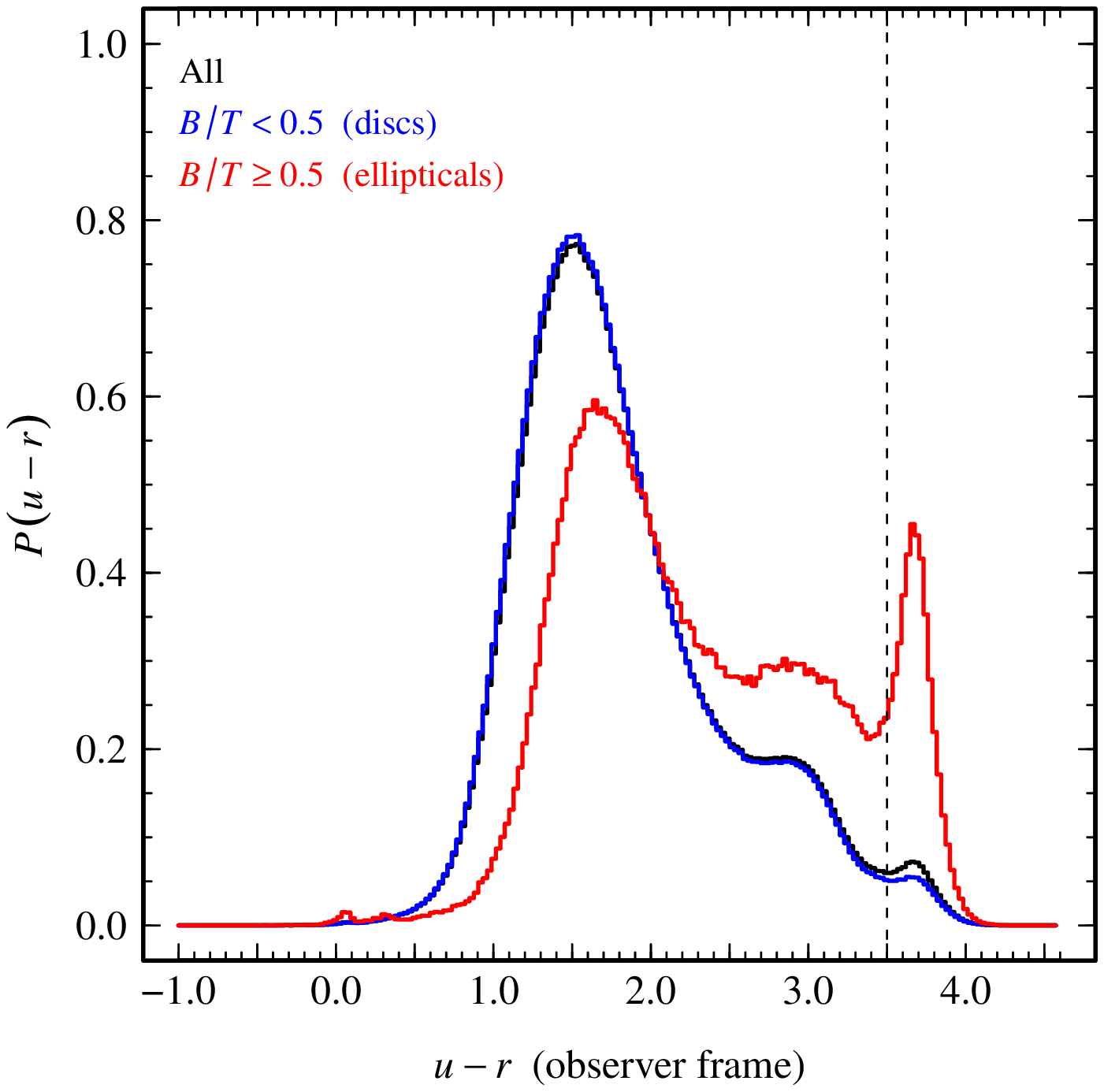}
  \caption{As Fig.~\ref{f:morphcolD}, but for the \mgal{}
    semi-analytic model; colours are therefore observer-frame.  The
    colour-cut is marked in the lower panel at $u-r = 3.5$.
}
  \label{f:morphcolM}
\end{figure}

When considering how the Split model operates when selecting systems
by galaxy colour, we need to examine the relationship between colour
and morphology.  Although they are classically seen to correlate well
\citep[e.g.][]{1961ApJS....5..233D, 1986ApJ...302..564S,
  2001AJ....122.1861S, 2004ApJ...608..752B}, both theoretical and
observational studies have shown the relationship to be not
straightforward, and based on detailed processes occurring during
galaxy formation and evolution \citep{2006MNRAS.365...11C,
  2007MNRAS.379..841B, 2007Ap.....50..273D, 2011MNRAS.413..101G}.  We
show the relationships between galaxy colour and morphology for our
models in Figs.~\ref{f:morphcolD} \&~\ref{f:morphcolM}.

In the \dgal{} model (Fig.~\ref{f:morphcolD}), it is clear that blue
galaxies are mainly discs, and disc galaxies are mainly
blue. Elliptical galaxies have a broader range of colours, albeit
dominated by red galaxies.  This means that, in the stacked halo
results, selecting just blue galaxies yields a much more circular
shape than selecting just red galaxies, although the mixing between
colour and morphology means that the situation in both cases is less
extreme than when selecting by morphology directly.

For the \mgal{} model (Fig.~\ref{f:morphcolM}), the population remains
dominated by disc galaxies when either blue or red galaxies are
selected, although it is to a lesser extent in the red
case. Similarly, both the disc galaxy and elliptical galaxy
populations are dominated by blue galaxies, with the elliptical galaxy
population having a significant red galaxy presence too.  This results
in a much smaller difference between the stacked halo shapes when
selecting just red and just blue galaxies, when compared to the
results from the \dgal{} model; in particular, the result for the red
population is significantly more circular.  However, the tight link
between blue and disc galaxies means that for both the \dgal{} and
\mgal{} model, selecting blue galaxies yields the more circular
stacked halo.

\subsection{Parallel \& Fitted alignment, and galaxy--halo correlations}
The results in Figs.~\ref{f:stcksDGals} \&~\ref{f:stcksMGals} for the
Parallel model (and to a lesser extent the Fitted model) also depend
on the colour and morphological selection, with red or elliptical
galaxies resulting in a more circular stacked halo than blue or disc
galaxies.  In this case, the difference is not due to the alignment
model, but an \emph{intrinsic} correlation between galaxy
colour/morphology and halo shape.

\begin{figure}
  \centering
  \includegraphics[width=\figw]{\fpath 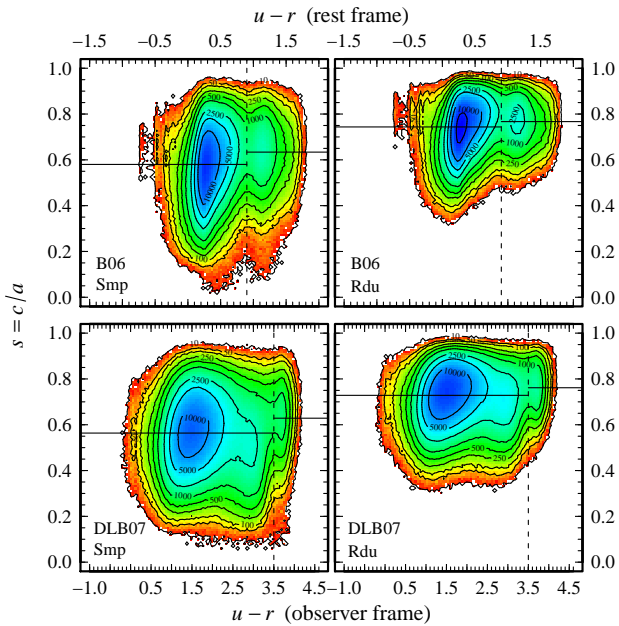}
  \caption{Joint distribution of 3-D halo shapes and the colours of
    their central galaxy.  Shapes are computed using the the $\Msmp$
    tensor (left panels) and the $\Mrdu$ tensor (right).  Top row:
    Results from the \dgal{} model, using rest-frame colours. Bottom
    row: Results from the \mgal{} model, using observer-frame colours.
    The vertical dashed line marks the red--blue division in both
    cases, and the horizontal solid lines mark the medians for red and
    blue galaxies separately.  
}
  \label{f:scol}
\end{figure}

\begin{figure}
  \centering
  \includegraphics[width=\figw]{\fpath 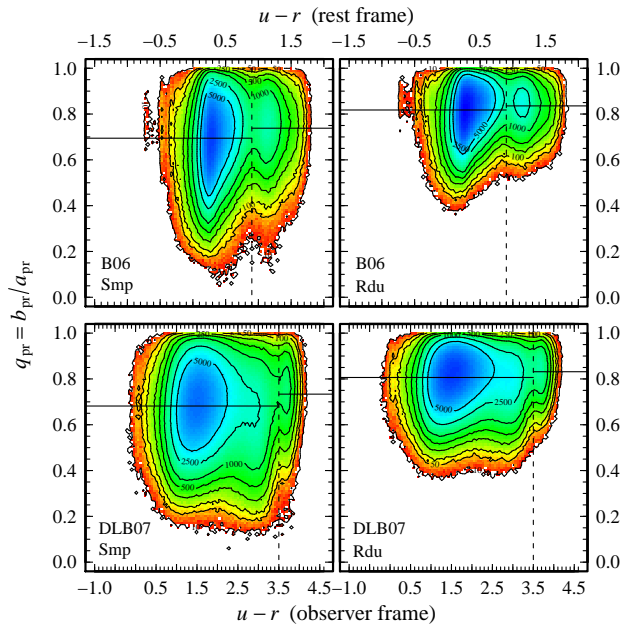}
  \caption{As Fig.~\ref{f:scol}, but using the projected halo shapes,
    assuming the Parallel alignment model. 
}
  \label{f:qcol}
\end{figure}

\begin{figure}
  \centering
  \includegraphics[width=\figw]{\fpath 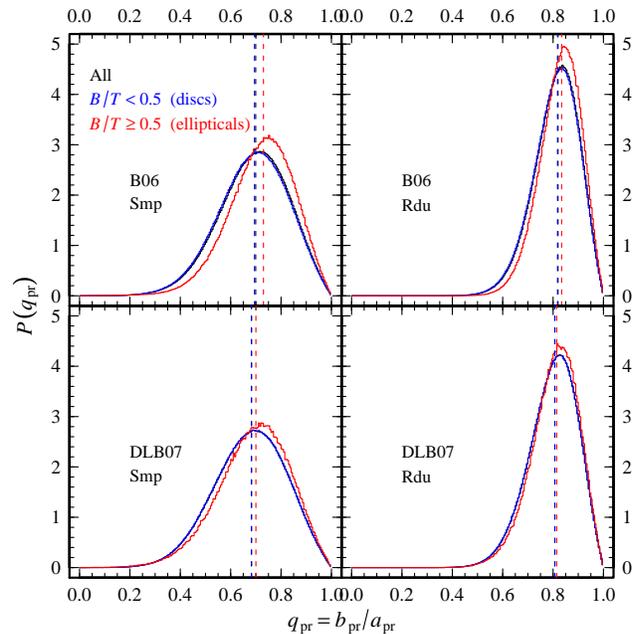}
  \caption{Histograms of projected halo shapes, assuming the Parallel
    alignment distribution, divided according to galaxy morphology:
    the distribution for elliptical galaxies is shown in red, and that
    of disc galaxies is shown in blue. The result from the full
    population is shown in black, but this closely matches that of the
    disc sample.  As in Figs.~\ref{f:scol} \& \ref{f:qcol}, we show
    the results from both semi-analytic galaxy models, and the $\Msmp$
    \& $\Mrdu$ shape measurement tensors.  Medians for each sample are
    shown by vertical dashed lines.  
}
  \label{f:qhistmorph}
\end{figure}

Figure~\ref{f:scol} shows the joint distribution of 3-D halo
sphericity $s$ from the $\Msmp$ and $\Mrdu$ tensors, against $u-r$
colour for the \dgal{} and \mgal{} semi-analytic models.  The
projected halo shapes in these cases, for the Parallel alignment
model, are shown against colour in Fig.~\ref{f:qcol}.  We can see that
the colour distribution is very broad for any given halo shape,
although the haloes of blue galaxies have a less-spherical median
shape than those of the red galaxies.  Similarly, we plot projected
axis ratio histograms in Fig.~\ref{f:qhistmorph}, cut by galaxy
morphology: ellpitical galaxies have slightly more circular projected
haloes in the median than disc galaxies.  It is important to note that
the median shape in a distribution is \emph{not} the same as the
stacked halo shape from the same sample of haloes, because the
stacking process weights haloes differently.

\begin{figure}
  \centering
  \includegraphics[width=\figwtwo]{\fpath 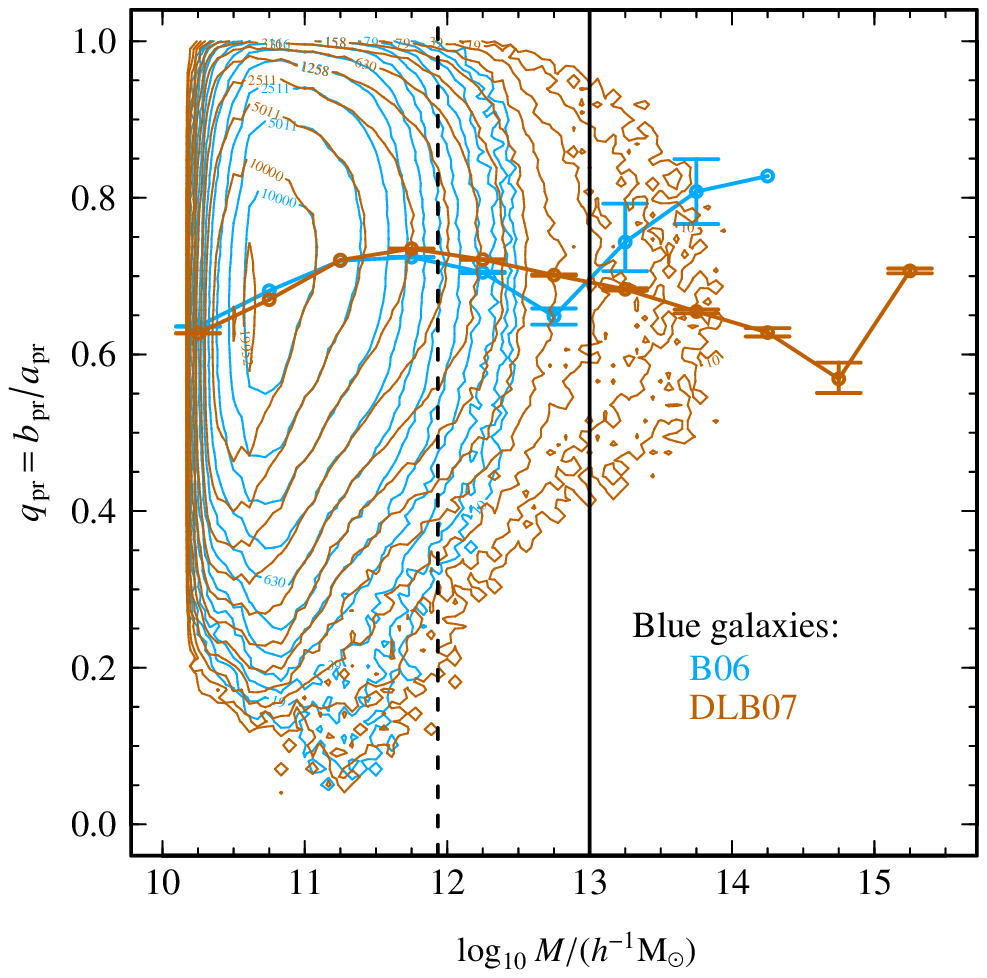}
  \includegraphics[width=\figwtwo]{\fpath 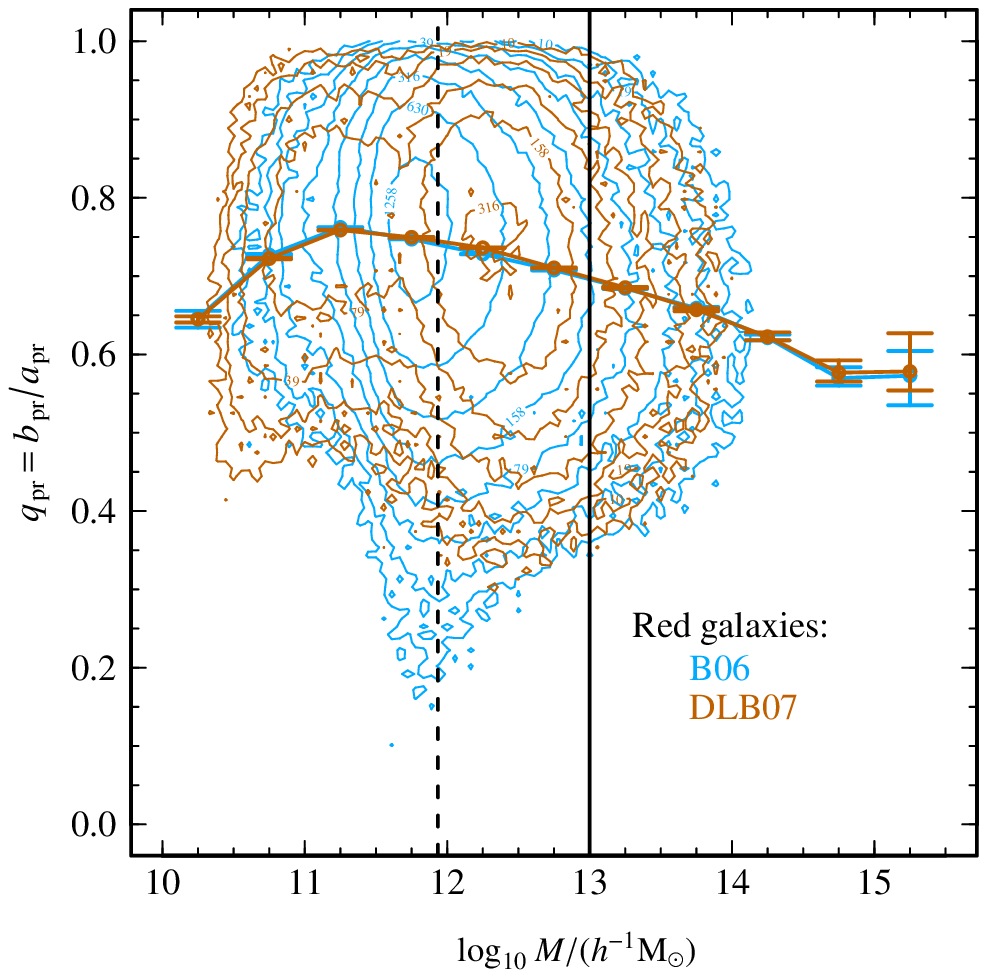}
  \caption{Joint distributions of projected halo axis ratio (assuming
    the Parallel distribution and using the $\Msmp$ tensor), with halo
    mass.  Each panel shows the results from both semi-analytic
    models, with the upper panel showing the blue population, and the
    lower panel showing the red population.  The solid line joins
    points giving the median in mass bins, with error bars given by
    the standard error on the median (equation~\ref{e:mederr}).  The
    same contour levels are used for both models in both panels.
}
  \label{f:qmasscols}
\end{figure}

While most of the results for Parallel alignment are very much
consistent between the \dgal{} and \mgal{} models, the projected
shapes for blue (and disc-dominated) galaxies using the $\Msmp$ tensor
are noticeably less circular for the \dgal{} model.  This is again due
to the $\Msmp$ shapes being dominated by the very largest haloes: In
the \dgal{} model, there are very few large haloes hosting blue
central galaxies (just $242$ systems with masses
$10^{12.5}$--$10^{13}\munit$), whereas with the \mgal{} model the blue
population extends to much higher masses (in the same mass bin, there
are $29\;905$ haloes).  We show the distributions of galaxies of
different colours, as functions of their parent halo mass, in
Fig.~\ref{f:qmasscols}.  The figure shows clearly how the highest-mass
haloes (up to $10^{13}\munit$) hosting blue galaxies in the \dgal{}
model have a lower median projected axis ratio than those in the
\mgal{} model.  Furthermore, despite the significant differences in
the distributions of red galaxies between the \dgal{} and \mgal{}
models, the medians as a function of mass are very similar.

It is important to note that the correlation between galaxy properties
and halo shape is relatively weak, and (as shown from the Monte Carlo
tests earlier, Fig.~\ref{f:MCstacksfixeds}) only plays a role when the
impact of galaxy--halo misalignment is strongly reduced.  Applying the
Fitted distribution provides an example of an intermediate case: while
there is still some variation between the results for red/elliptical
and blue/disc galaxies, the circularising effect of galaxy--halo
misalignment means that these differences are negligable. Indeed, the
stacked halo results are sufficiently circular under the Fitted
distribution ($q_\pr \ga 0.95$) that it would be extremely difficult
for a weak lensing study to measure any significant ellipticity.



\section{Discussion and conclusions}\label{s:conc}
\cite{2000ApJ...538L.113N} and \cite{2000astro.ph..6281B} described a
technique for measuring halo ellipticity using the azimuthal variation
in the tangential shear signal.  Since this is a weak variation on top
of the already weak shear signal, any measurement would require very
large numbers of galaxies, from large surveys.  In this paper, we have
shown quantitatively, for a range of possible models, that a
measurement of anisotropy would be extremely difficult indeed.  For
example, using a model based on current hydrodynamic galaxy formation
simulations yields stacked halo axis ratios of $q_\pr \ga 0.95$, which
would be extremely difficult to observe.

The main problem, as expected, is the galaxy--halo misalignment.  Any
intrinsic correlations between halo shape and galaxy properties are
only relevant in the case of unrealistically perfect alignment.  While
this can be seen in our main results from using the Millennium
Simulation, it is very clearly demonstrated in just using simple Monte
Carlo tests.

It is important to note, when considering observations related to
galaxy--halo alignment, that the \emph{physically} relevant angle
$\theta$, as measured from simulations, is the alignment in 3-D.  Due
to the random orientations of the other angles in the system (galaxy
minor axis polar angle $\phi$, galaxy orientation $\xi$ and image
plane inclination $\zeta$), even parallel alignment with $\theta=0$
does not necessarily lead to perfect alignment of the
\emph{observationally} relevant angle, between the projected axes.

In both of the semi-analytic galaxy formation models we test here, we
find that blue or disc-dominated galaxies tend to reside in
less-spherical haloes.  However, if we assume that elliptical galaxies
are aligned more closely to their halo than disc galaxies (following
e.g. \citealt{2004MNRAS.347..895H}), then it is selecting red or
elliptical galaxies that yields a more elliptical stacked halo in
projection.  In our work, this of course occurs by construction, and
further work on galaxy--halo alignment in simulations is required to
see how plausible this is in practice.  However, the alignment
distributions we use span the range of possibilities, and, in
conjunction with our Monte Carlo tests and the colour/morphology
distributions of modelled galaxies, the result of any arbitrary
alignment distribution can be estimated.

The work presented in this paper can be seen as a dark
matter-theoretical counterpart to the paper of
\cite{2010MNRAS.407..891H}, who performed a thorough study of
difficulties with measuring anisotropic shear from a
lensing-theoretical standpoint.  Our distributions for the projected
halo axis ratio (Figs.~\ref{f:scol}--\ref{f:qmasscols}) derive from
more complex modelling for example, and show the possible variability
due to galaxy type, but \cite{2010MNRAS.407..891H} take a given
distribution through a Monte Carlo lensing process, demonstrating that
multiple deflections of background galaxies can \emph{also} have a
catastrophic effect on the anisotropic shear signal.

Our results may lead one to wonder how it was that both \cite{hyg2004}
and \cite{parker2007} managed to claim a measurement of
 halo ellipticity.  It should
first be mentioned that their detections are relatively weak anyway,
\pbnew{with their halo ellipticity measurements being} at the
$1$--$2\sigma$ level \pbnew{(although \citealt{hyg2004} rule out
  spherical haloes at the $3\sigma$ level)}.  Possible sources of the
discrepancy include underestimation of systematic errors in the
modelling and interpretation of the data, not helped by the lack of
redshift and colour/morphology data in both of these studies.
\cite{mandelbaum2006} discuss in some detail the discrepancy between
their non-detection and the results of \cite{hyg2004}.  It is clear
that, regardless of the physics-related problems highlighted in the
present work, controlling systematics \pbnew{in}
observational studies \pbnew{such as these} is very difficult.

We have already mentioned the difficulty in using this method as a way
of distinguishing between \lcdm{} and modified gravity theories.  The
basic idea -- \lcdm{} predicts non-spherical haloes, but modifications
to gravity without dark matter predict spherical symmetry -- is based
on a na\"ive understanding of practical issues in both the \lcdm{} and
modified gravity case.  With \lcdm{}, systematic effects in the
lensing and stacking procedure can easily render the net signal
isotropic, and as we have shown, our poor knowledge of the
relationship (and in particular the alignment) between galaxies and
their parent haloes provides much of the uncertainty.  In the case of
galaxies in modified gravity theories, all the aforementioned problems
with our understanding of the baryonic physics still apply, but in the
context of gravity laws that are more complex and less well
understood.  There has been some simulation work with gas dynamics in
MOND \citep{2008A&A...483..719T}, but no full simulation of galaxy
formation in a cosmological context, with any star formation or
feedback (simulations in \lcdm{} have shown how significant an impact
these processes have on the resulting galaxies).  In the context of
STVG, the gravity law is more difficult to work with, and numerical
simulations are still in their infancy \citep[see
e.g.][]{2010arXiv1005.2685M}.  Thus, we do not believe that
statistical analysis of stacked, projected lens galaxies can be used
to discriminate between \lcdm{} and alternative theories, simply
because we lack robust predictions from either case.

Consequently, in this paper, we have not gone as far as to make a
prediction for observations, as the theoretical uncertainty is still
too large.  In the future however, if the galaxy formation models
reach better convergence and can offer statistical predictions of
galaxy--halo alignment, then a study such as ours could be advanced
further to make such an observational prediction.  In that case,
certain other effects would need to be taken in to account.  We have
been able to neglect these here, as they are all secondary to the main
misalignment difficulty.

Firstly, when computing shapes from simulations to compare with
observations, then it would be more appropriate to use the mass within
a given radius.  This is (arguably) not the same as the shape of the
halo, which is a dynamically relaxed physical structure, rather than a
geometrically-defined overdense region.  However, an observation such
as this has no practical way of accessing the dynamical information
necessary to define a halo, and so there is no need to do so in
simulations for this purpose either\footnote{It should be noted that
  in analyses of simulations alone, the distinct ideas of measuring
  the shape of an overdensity contour colocated on the same density
  peak as a halo, and the shape of the dynamically-defined halo itself
  are often conflated.}.  Ideally, the mass distribution as a function
of radius would be generated from the simulations, as it has been
shown that halo ellipticity is not constant with radius.  The
distinction between central and satellite galaxies could also be
relaxed, and the shape of the mass distribution around each (lens)
galaxy could be computed, allowing for selection criteria that more
closely match those in observational studies. \pbnew{(The problem of which
galaxy is at the centre of a halo is not limited to simulations; for
example,  \cite{2011MNRAS.410..417S} have shown that the brightest
galaxies in haloes are often not in the centre, and this should be
taken into account when simulating observations.)}

When considering the inner halo shape however, it becomes vital to
consider the impact of baryonic processes.  While dark matter-only
haloes are triaxial with a tendency for prolateness, becoming more
prolate towards the centre \citep[e.g.][]{bett2007,
  2007MNRAS.377...50H}, haloes that have had a galaxy form in the
centre are overall more spherical, with a tendency towards oblateness
\citep{2004ApJ...611L..73K, 2010ApJ...720L..62K, 2005ApJ...627L..17B,
  2006ApJ...648..807B, 2006PhRvD..74l3522G, 2008ApJ...681.1076D,
  2010MNRAS.406..922T, 2010MNRAS.406.2386M, 2010MNRAS.407..435A,
  2011ApJ...734...93L}.  This is likely to make halo shapes more
difficult to measure.  On the other hand, strong lensing studies have
suggested that mass and light are well aligned in the inner regions of
the halo \citep{2002sgdh.conf...62K,LensNotesSL, 2008MNRAS.391..653M}.

It would be important to measure the alignment distribution from a
statistically large sample of objects, and over a range of time steps.
It is known that both halo and galaxy orientations vary in time
\emph{even outside major mergers} \citep[e.g.][]{2009MNRAS.396..696S,
  2009ApJ...702.1250R, 2010AIPC.1240..403B, 2011arXiv1104.0935B}, so
the relative orientations of a few galaxies and haloes at a single
redshift might not be at all robust, regardless of how well-resolved
they are spatially, or how realistic the baryonic physics in the
simulation is.

Eventually, realistic mock-observations would need to be produced,
using ray tracing through the simulation
\citep[e.g.][]{2009A&A...499...31H} so that a realistic background
source population and the effect of multiple deflections are included,
as it has been shown that these have a significant impact on shear
measurements \citep{2010ApJ...713..603B, 2010MNRAS.407..891H,
  2011MNRAS.412.2095H}.

While our results do not seem to give much cause for optimism in
measuring shapes using weak lensing, it should be pointed out that
\pbnew{we are only considering one method, for the shape distribution
  of non-cluster haloes. Many} other methods of
measuring halo shapes are possible, and indeed are actively persued.
\pbnew{Furthermore, the ideal test of \lcdm{} is to measure the
  inceasing asphericity of haloes with increasing mass, and thus the
  shapes of cluster haloes are particularly important.}
\cite{2009ApJ...695.1446E} applied essentially the same technique as
\cite{2000ApJ...538L.113N}, but on clusters rather than field
galaxies.  They managed to measure a projected halo axis ratio of
$q_\pr = 0.48^{+0.14}_{0.09}$ with $1\sigma$ errors, ruling out a
circular shape at $99.6\%$ confidence.  Using clusters has the
practical advantage that the cluster member galaxies can be used for
alignment.  Much work has been done on the alignment of the satellite
galaxy distribution, both for measuring cluster halo shapes and as
another test of dark matter, observationally and in simulations
\citep[e.g.][and references therein]{2007ApJ...671.1135K,
  2008ApJ...675..146F, 2009MNRAS.399..550L, 2010MNRAS.405.1119K,
  2006MNRAS.369.1293Y, 2008MNRAS.385.1511W, 2010ApJ...709.1321A}.
\cite{2005ApJ...618....1H} also investigated the distribution of
projected halo shapes, for simulated cluster-mass haloes as a function
of redshift.  Other cluster-based lensing methods attempt to map the
shape directly \citep[e.g.][in the latter case finding $q_\pr=0.54\pm
0.04$ at $1\sigma$]{2004ApJ...613...95C, 2008MNRAS.383..857F,
  2010MNRAS.405.2215O, 2010ApJ...721..124D}, or use Markov Chain Monte
Carlo methods to fit triaxial models \citep{2009MNRAS.393.1235C}.
Lensing flexion has recently been proposed as another method for
studying galaxy-scale haloes \citep{2011A&A...528A..52E,
  2011MNRAS.417.2197E}.  Non-lensing methods for studying halo shape
include studying the distribution of H\,\textsc{i} in disk galaxies
\citep[e.g.][and references therein]{2008ApJ...685..254B,
  2010A&A...515A..63O}.

In this paper, we have presented a quantitative analysis of the
 \pbnew{impact} of galaxy--halo misalignment on the
possiblity of measuring halo shapes via weak lensing in stacked
images.  We have tested a  \pbnew{series} of alignment
models, spanning the range from perfect alignment (in 3-D) to
uniformly-distributed alignment.  As intermediate models, we included
a fit to recent hydrodynamic simulations of galaxy formation, and a
distribution that explicitly differentiates between galaxy
mophologies.  Our results have shown that, for there to be a
reasonable possiblity for shapes to be measured, a significant
fraction of the lens galaxies must have close to perfect alignment,
which seems physically implausible.  Using simple Monte Carlo models,
we have quantified how well-aligned the galaxies have to be in their
haloes before the intrinsic shape distribution becomes measurable.
For our results using the Millennium Simulation, we have also tested
the impact of using different models of galaxy formation, and
different ways of measuring haloes in simulations.  These illustrate
some of the difficulties in applying results from current simulations
directly to models: there simply \pbnew{is} not a single
robust quantitative prediction from \lcdm{} for halo shape
measurements using this method.  Since the same is true for
alternative theories without dark matter, this method cannot yet be
used to falsify one or the other.

\section*{Acknowledgements}
The author thanks Peter Schneider and Philippe Heraudeau for helpful
discussions, and Alis Deason for providing the GIMIC galaxy--halo
alignment data. This work was supported by the Deutsche
Forschungsgemeinschaft under the project SCHN 342/7--1 in the
framework of the Priority Programme SPP-1177, and the Initiative and
Networking Fund of the Helmholtz Association, contract HA-101
(``Physics at the Terascale'').  The simulations used in this paper
were carried out as part of the programme of the Virgo Consortium on
the Regatta supercomputer of the Computing Centre of the
Max-Planck-Society in Garching, and the Cosmology Machine
supercomputer at the Institute for Computational Cosmology, Durham.
The Cosmology Machine is part of the DiRAC Facility jointly funded by
STFC, the Large Facilities Capital Fund of BIS, and Durham University.
The Millennium Simulation databases used in this paper and the web
application providing online access to them were constructed as part
of the activities of the German Astrophysical Virtual Observatory.


%
\bibliographystyle{mn2eimproved}
\bibliography{wlhaloshapes}

\appendix

\section{The Azimuthally-averaged Fisher distribution}\label{s:fishdistro}
The probability distribution we use in section~\ref{s:fitdistro} to
define our Fitted alignment distribution is based on the
\cite{fisher53} distribution,
\begin{equation}
  \label{e:Fisher}
  P_\rmn{F}(\vv{v};\vzero,\kappa) 
  = \frac{\kappa}{\sinh \kappa} \exp\left(\kappa\;\vv{v}\cdot\vzero\right),
\end{equation}
where the probability density function (PDF) is given in terms of the
unit vector random variable $\vv{v}$, the mean direction unit vector
$\vzero$, and the concentration $\kappa$; the latter is often written
in terms of the width of the distribution $\sigma$ through $\kappa =
1/\sigma^2$.  This is the 3-D case of the more general von
Mises--Fisher family of distributions, and is often used as more
mathematically tractable approximation to a wrapped Normal
distribution (see \citealt{MardiaJupp} for more details).

If we write our vectors in a cartesian basis in terms of spherical
polar coordinates, and (for our case) take the random variable
$\vv{v}$ to be the galaxy axis $\vcgal$, oriented with respect to the
halo vector $\vh$ located on the $z$-axis, we can write
\begin{equation}
  \vv{v}=\vcgal = \cvv{\sin\theta \cos\phi}{\sin\theta \sin\phi}{\cos\theta},
  \;
  \vzero = \cvv{\sin\theta_0 \cos\phi_0}{\sin\theta_0 \sin\phi_0}{\cos\theta_0}.
\end{equation}
Normalisation of the PDF is over the surface of the unit sphere
($\int_\rmn{S} \dd\Omega = \int_0^{2\pi}\int_0^\pi \sin\theta
\dd\theta \dd\phi = 4\pi$), so we can write the PDF in terms of
$\theta$ and $\phi$ as:
\begin{equation}
   P_\rmn{F}(\theta,\phi) = \frac{\kappa}{4\pi \sinh \kappa} 
    \rmn{e}^{\kappa[\cos\theta\cos\theta_0 + \sin\theta\sin\theta_0\cos(\phi-\phi_0)]}
   \sin\theta,
 \end{equation}
such that the normalisation integral is 
\begin{equation}\label{FishNorm}
  \int_0^{\pi} \int_0^{2\pi} 
  P_\rmn{F}(\theta,\phi) 
  \; \dd\phi \; \dd\theta = 1.
\end{equation}

However, in our case, we are only interested in the angle $\theta$
between our two vectors, so we have to integrate the Fisher
distribution over all values of the azimuthal angle $\phi$.  The
$\phi$-integral is in fact related to the zeroth-order modified Bessel
function of the first kind, $I_0(x)$:
\begin{equation}
  \label{e:phiintegral}
  \int_0^{2\pi}  \exp\left[\kappa\sin\theta\sin\theta_0\cos(\phi-\phi_0)\right]
   \dd\phi
    = 2\pi I_0(\kappa\sin\theta\sin\theta_0).
\end{equation}
so we can write our azimuthally-averaged Fisher distribution as 
\begin{equation}
  P(\theta) = 
  \frac{\kappa}{2 \sinh \kappa}\;
  I_0(\kappa\sin\theta\sin\theta_0)\;
  \exp\left(\kappa\cos\theta\cos\theta_0\right)\; \sin\theta.
\end{equation}
To aid comparison with the uniform distribution, we shall actually
normalise in $\cos\theta$ instead of $\theta$, so the final PDF
that we use is:
\begin{equation}
  P(\cos\theta) = 
  \frac{\kappa}{2 \sinh \kappa} \;
  I_0(\kappa\sin\theta\sin\theta_0) \;
  \exp\left( \kappa \cos\theta \cos\theta_0 \right),
\end{equation}
which is equation~(\ref{e:azavfish}).

Note that for large widths ($\sigma \ga 10$), the PDF tends to the
uniform distribution, and for narrow widths the PDF tends to a delta
function spike at $\theta_0$.  

The routine we use to sample from this distribution is based on that
given in the \textsc{Prob}
library\footnote{\url{http://people.sc.fsu.edu/~jburkardt/cpp_src/prob/prob.html}}
of John Burkardt, which in turn is based on
\cite{FisherLewisEmbleton2003}.

\section{Rotation and projection}\label{s:rotmats}
Here, for completeness, we give further details of our method for
defining the rotations and projection involved in our galaxy--halo
alignment model described in section~\ref{s:galrot}.  Consider a
reference frame $S$, and a second frame $S'$ that is a rotation of
$S$.  If we write the cartesian basis vectors of $S'$ in terms of
those of $S$, e.g. $\ux = (\hat{x}_1,\hat{x}_2,\hat{x}_3)^\T$, then we
can simply write down the rotation matrix that transforms from $S$ to
$S'$:
\begin{equation}
\mat{R} = \left(\begin{array}{lll}
    \hat{x}_1 & \hat{x}_2 & \hat{x}_3 \\
    \hat{y}_1 & \hat{y}_2 & \hat{y}_3 \\
    \hat{z}_1 & \hat{z}_2 & \hat{z}_3 
  \end{array}\right)
\end{equation}

Thus, given the basis vectors for the halo coordinate frame in terms
of the simulation coordinates, given in equation~(\ref{e:hbasis}), we
can write down the rotation matrix for transforming from the
simulation coordinates into these halo-based coordinates:
\begin{equation}\label{e:Rhalo}
  \mat{R}_\hlo = \left(\begin{array}{lll}
      \phantom{-}\cos\theta_\hlo \cos\phi_\hlo & \cos\theta_\hlo \sin\phi_\hlo & -\sin\theta_\hlo \\
                -\sin\phi_\hlo               & \cos \phi_\hlo             & \phantom{-}0 \\
      \phantom{-}\sin\theta_\hlo \cos\phi_\hlo & \sin\theta_\hlo \sin\phi_\hlo & \phantom{-}\cos\theta_\hlo \\
    \end{array}\right)
\end{equation}

In this halo frame, we define the orientation of the galaxy by first
specifying a minor axis vector direction given by $\theta$ and $\phi$.
We can thus rotate into a similar galaxy-vector-based frame using a
matrix $\mat{R}_\gal$ identical in form to equation~(\ref{e:Rhalo}).
However, the orientation of the galaxy about its minor axis must also
be specified, by a further rotation by the angle $\xi$.  Defining
$\xi$ such that setting $\xi=0$ makes $\mat{R}_\gal$ have the same
form as $\mat{R}_\hlo$, means that we can write down the full rotation
matrix as
\begin{equation}\label{e:Rgal}
  \mat{R}_\gal = \left(\begin{array}{lll}
\phantom{-}c_\theta c_\phi c_\xi -s_\phi s_\xi & \phantom{-}c_\theta s_\phi c_\xi +c_\phi s_\xi &           -s_\theta c_\xi \\
          -c_\theta c_\phi s_\xi -s_\phi c_\xi &           -c_\theta s_\phi s_\xi +c_\phi c_\xi & \phantom{-}s_\theta s_\xi \\
\phantom{-}s_\theta c_\phi                         & \phantom{-}s_\theta s_\phi                        & \phantom{-}c_\theta \\
\end{array}\right).
\end{equation}
where we have used $c_X$ and $s_X$ as shorthands for $\cos X$ and
$\sin X$, for brevity.

The image plane (section~\ref{s:imgplane}) is based on a projection of
the galaxy major and minor axes, corresponding to the image $x$ and
$y$ axes respectively, but also allowing for a rotation of $\zeta$
about the major axis.  This means that, to rotate into the image frame
from the galaxy frame, we use
\begin{equation}  \label{e:Rimg}
  \mat{R}_\img = \left(\begin{array}{lll}
      1 & \phantom{-}0 & \phantom{-}0\\
      0 & -\sin\zeta &  \phantom{-}\cos\zeta \\
      0 & -\cos\zeta & -\sin\zeta \\
    \end{array}\right).
\end{equation}

Putting these together, a vector in the simulation frame
$\vv{p}^\rmn{sim}$ can be transformed into the image frame simply by
\begin{equation}
\vv{p}^\img = \mat{R}_\img \mat{R}_\gal \mat{R}_\hlo \vv{p}^\rmn{sim}.
\end{equation}
In practice, we will have the halo mass distribution matrix $\mat{M}$
(see section~\ref{s:shapes}), measured in the simulation frame.  We
therefore transform this into the image frame by
\begin{equation}
  \mat{M}^\img = \left(\mat{R}_\img \mat{R}_\gal \mat{R}_\hlo \right)
  \mat{M} \left(\mat{R}_\img \mat{R}_\gal \mat{R}_\hlo \right)^{-1}.
\end{equation}
The projected mass distribution in the image plane is then the
top-left $2\times 2$ submatrix of $\mat{M}^\img$.  The eigenvalues and
eigenvectors of this can then easily be found, with the axis lengths
of the projected halo ellipse being given by the square root of the
eigenvalues.

\section{Results from other redshifts}\label{s:zresults}
We show here stacked projected halo shapes at different redshifts.
They show essentially the same dependences on halo and galaxy
properties as have already been illustrated, but nevertheless give an
idea of another parameter that can have a significant quantitative
impact on the results.

\begin{figure}
  \centering\includegraphics[width=\figw]{\fpath 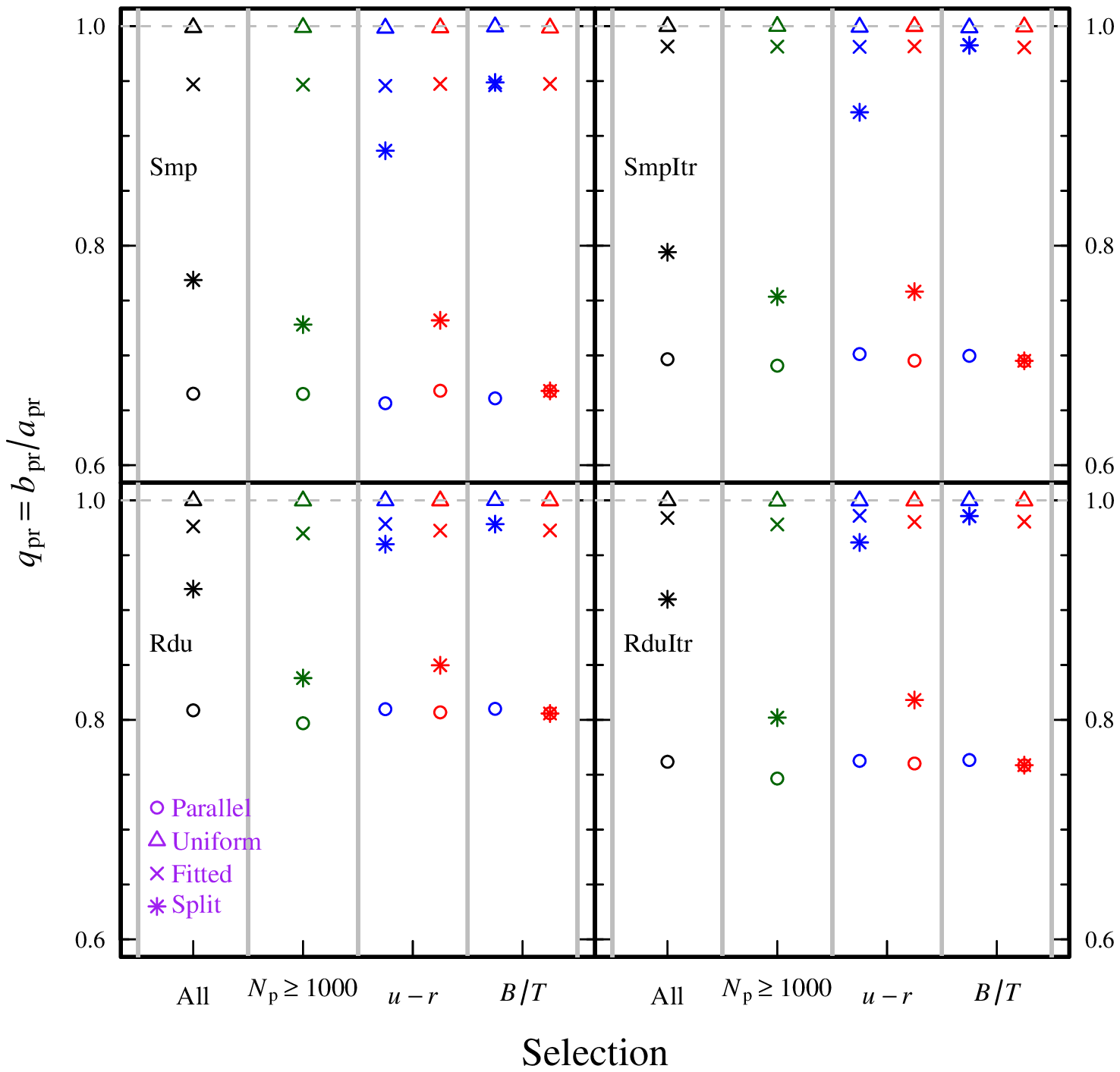}
  \caption{As Fig.~\ref{f:stcksDGals} (i.e. using \dgal{}), but using
    redshift $z\simeq 0.5$.
}
  \label{f:stcksDGalsHiz}
\end{figure}

\begin{figure}
  \centering\includegraphics[width=\figw]{\fpath 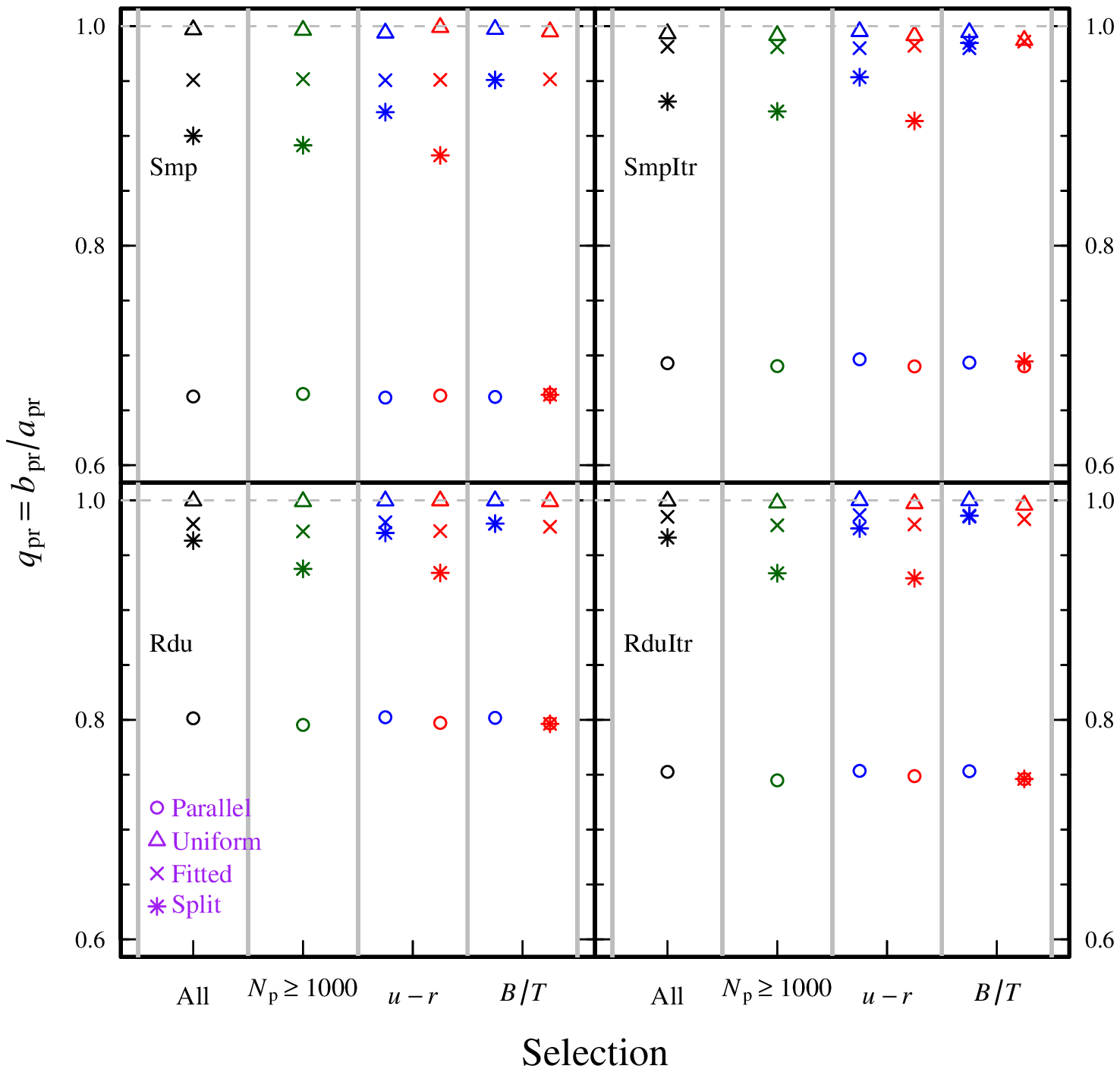}
  \caption{As Fig.~\ref{f:stcksMGals} (i.e. using \mgal{}), but using
    redshift $z\simeq 0.5$.
}
  \label{f:stcksMGalsHiz}
\end{figure}

We choose $z\simeq 0.50$ (MS snapshot 48) as a higher redshift for
analysis.  The results for the \dgal{} and \mgal{} models are shown in
Figs.~\ref{f:stcksDGalsHiz} and \ref{f:stcksMGalsHiz} respectively.
There is very little change from the $z\simeq 0.32$ results in
Figs.~\ref{f:stcksDGals} \& \ref{f:stcksMGals}.  When the Split
distribution is used, the results for ``All'' haloes are noticably
less circular, implying a greater proportion of elliptical galaxies.
This might initially be seem to contradict the findings of
\cite{2009MNRAS.396.1972P}: they showed that, at higher redshifts
there should be more disc galaxies, and fewer ellipticals.  However,
they found that this is strongest for the \mgal{} model (where we see
less change) and very weak for \dgal{} (where we see the largest
difference).  The key is that while \cite{2009MNRAS.396.1972P} select
bright galaxies at each redshift, selecting by $K$-band absolute
magnitude $M_K - 5\log_{10} h < -22.17$, we are using a deeper cut in
apparent magnitude: at this redshift, we are selecting galaxies
brighter than $M_r - 5\log_{10}h = -17.3$.  Thus, we are sampling more
low-mass systems than \cite{2009MNRAS.396.1972P} (at all redshifts),
but fewer than we were at $z\simeq 0.32$. Our sample has
proportionally more higher-mass systems (the biggest difference in
shape is with $\Msmp$), which tend to host elliptical galaxies.  The
changes among the bright galaxy population seen in
\cite{2009MNRAS.396.1972P} are secondary to the overall change in
galaxy demographic at higher redshift.

\begin{figure}
  \centering\includegraphics[width=\figw]{\fpath  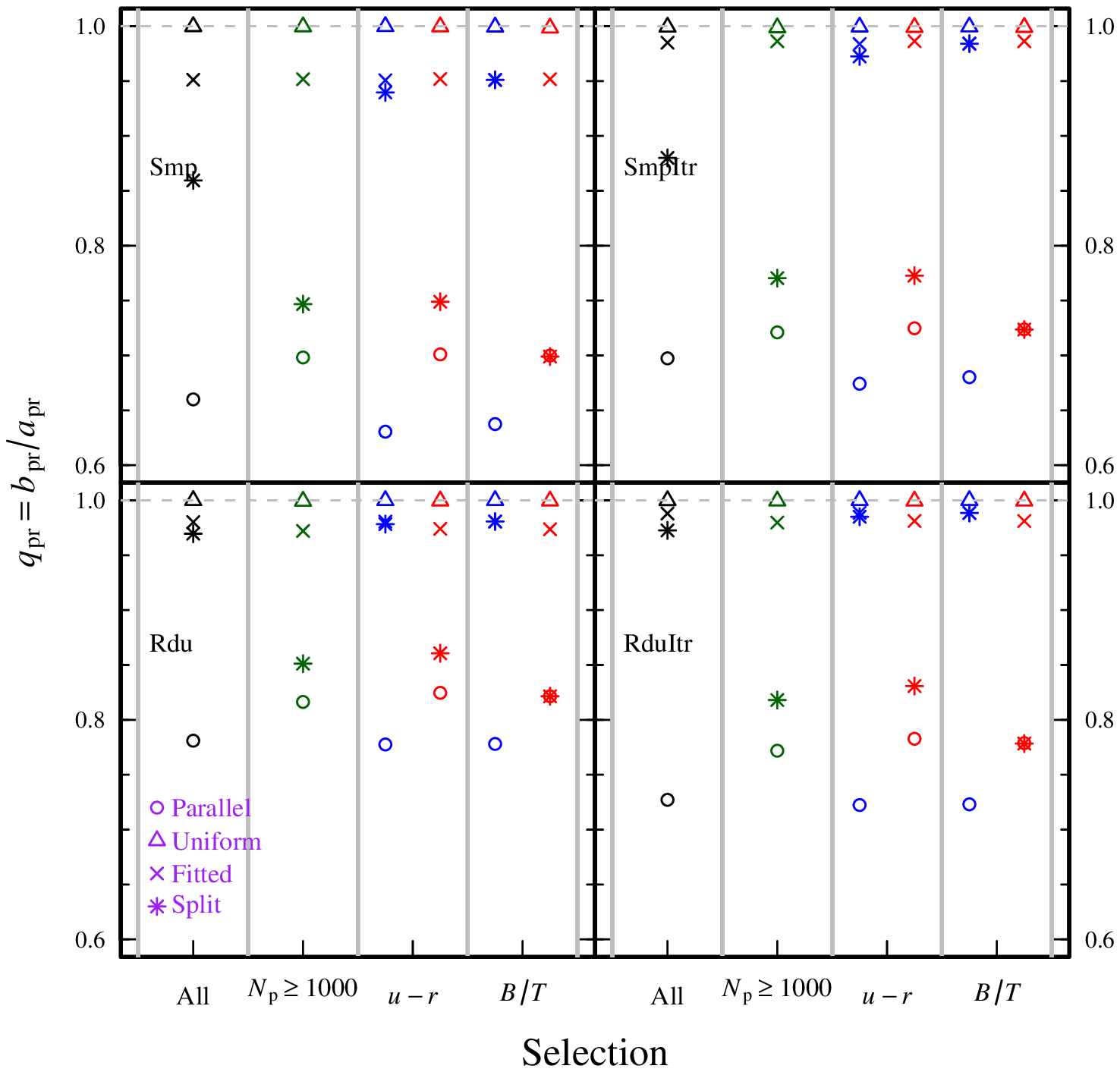}
  \caption{As Fig.~\ref{f:stcksDGals} (i.e. using \dgal{}), but using
    redshift $z\simeq 0.17$.
}
  \label{f:stcksDGalsLoz}
\end{figure}

\begin{figure}
  \centering\includegraphics[width=\figw]{\fpath 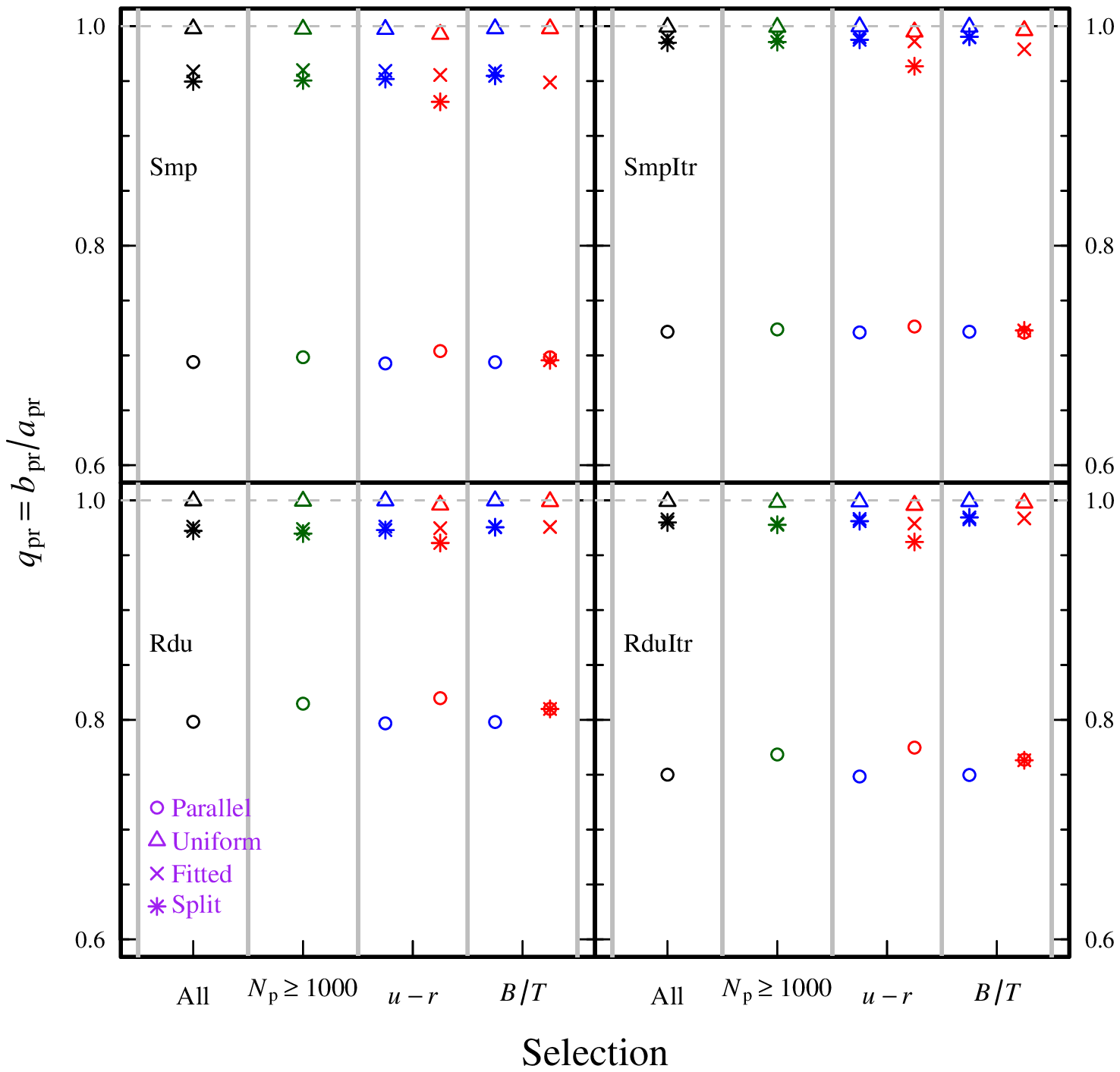}
  \caption{As Fig.~\ref{f:stcksMGals} (i.e. using \mgal{}), but using
    redshift $z\simeq 0.17$.
}
  \label{f:stcksMGalsLoz}
\end{figure}

Figs.~\ref{f:stcksDGalsLoz} \&~\ref{f:stcksMGalsLoz} show our results
at a lower redshift, $z\simeq 0.17$ (MS snapshot 56; here, our
selection cut is at $M_r - 5\log_{10}h = -14.6$).  In this case, the
results from the \dgal{} model show a noticably less circular shape
for blue and disc galaxies.  Since this is only significant for the
Parallel alignment model, it must be due to a greater \emph{intrinsic}
correlation between elliptical haloes and blue/disc-dominated galaxies
at this redshift.  Furthermore, the effect is not apparent in the
\mgal{} model.  The colour distribution in \mgal{} is much broader,
particularly for blue galaxies (see Fig.~\ref{f:coldistros}), such
that a colour cut is no longer an efficient way of selecting the more
aspherical haloes.

Taken together, these results emphasise those from the main body of
the paper.  Under the alignment model designed to fit recent galaxy
formation simulations, the alignment is sufficiently poor that very
little changes the result. However, if there is a significant
population with very good alignement, then the resulting stacked shape
will depend sensitively on the mass, shape, colour and morphology
distribution of the galaxy--halo systems; all of which depends on
redshift, and, at present, the galaxy model used.


\label{lastpage}
\end{document}